# Pseudo-Haptics Survey: Human-Computer Interaction in Extended Reality & Teleoperation

Rui Xavier[1,2], José Luís Silva[4], Rodrigo Ventura[1,2] (Member IEEE), and Joaquim Jorge[1,3] (Senior Member, IEEE)

[1] Instituto Superior Técnico (IST), University of Lisbon, Portugal
[2] Institute for Systems and Robotics (ISR-Lisboa), Portugal
[3] INESC-ID, Lisbon, Portugal
[4] ITI/LARSyS, Instituto Universitário de Lisboa (ISCTE-IUL), Lisboa, Portugal

Corresponding author(s): rui.dias.xavier@gmail.com

This work was supported by LARSyS funding (DOI:10.54499/LA/P/0083/2020, 10.54499/UIDP/50009/2020, and 10.54499/UIDB/50009/2020), INESC-ID Funding UIDB/50021/2020 (DOI:10.54499/UIDB/50021/2020) and 2022.09212.PTDC. This paper was written under the auspices of the UNESCO Chair on AI&XR.

**ABSTRACT** Pseudo-haptic techniques are becoming increasingly popular in human-computer interaction. They replicate haptic sensations by leveraging primarily visual feedback rather than mechanical actuators. These techniques bridge the gap between the real and virtual worlds by exploring the brain's ability to integrate visual and haptic information. One of the many advantages of pseudo-haptic techniques is that they are cost-effective, portable, and flexible. They eliminate the need for direct attachment of haptic devices to the body, which can be heavy and large and require a lot of power and maintenance. Recent research has focused on applying these techniques to extended reality and mid-air interactions. To better understand the potential of pseudo-haptic techniques, the authors developed a novel taxonomy encompassing tactile feedback, kinesthetic feedback, and combined categories in multimodal approaches, ground not covered by previous surveys. This survey highlights multimodal strategies and potential avenues for future studies, particularly regarding integrating these techniques into extended reality and collaborative virtual environments.

## I. INTRODUCTION

Pseudo-haptics is a sensation illusion perceived by playing with multimodal feedback, primarily visual, and exploring the brain's capabilities and limitations. By introducing subtle nuances of pseudo-haptic techniques (PHTs) mapped to a user's action, it is possible to simulate virtual tactile and kinesthetic sensations through multimodal simulated sensory effects, such as visual and auditory effects, or embodied metaphors, induced without needing a haptic device attached or applied to the body [1]. For instance, one can feel a virtual object's weight [2] or the pressure of a virtual flowing river [3]. The term pseudo-haptics seems to be contradictory, since "haptic", derived from the Greek word "haptesthai," meaning "to touch," refers to the science of feeling and manipulating through touch [4]. Haptic perception occurs when manipulating real objects. Therefore, haptics research aims to simulate similar perceptions (e.g., weight, heat, force, friction, or roughness) by providing feedback that can be kinesthetic (force felt by internal sensors in muscles, joint angles of arms, hand, wrist, fingers, etc.) or cutaneous (through sensors in human skin). Haptic feedback is crucial for remote interactions (e.g., in robotic teleoperation) and for enhancing immersive experiences in virtual environments. This feedback is used in desktop displays, touchscreen interfaces, or more complex extended reality (XR) environments, including augmented reality (AR), virtual reality (VR) with head-mounted displays (HMD), or settings that combine virtual and real elements, such as mixed reality (MR). In such environments, the sense of presence results from multi-sensory (visual, auditory, tactile, and olfactory) stimulation. A significant challenge is the lack of haptic feedback and naturalness, especially for mid-air gestural interactions where a person is not holding a controller device. Indeed, providing force feedback is usually associated with haptic devices, which extend the traditional display-based human-computer interfaces (HCI) by exploiting the sense of touch. However, these devices restrict user motion and introduce portability constraints, even with advancements with untethered wearable haptic controllers used in XR. Such devices may also generate unintended noise, require frequent maintenance, and have complicated interfaces.

Moreover, haptic devices retain the same physical fixed shape and appearance regardless of function, although some research focuses on dynamically changing the device's shapes [5]. One key advantage of pseudo-haptics is their device-free nature, making interfaces more flexible, allowing a broader range of user motion, and improving accuracy and portability by enabling dynamic modifications to the virtual properties of objects, surfaces, or avatars. PHTs offer cost advantages compared to haptic device hardware solutions and address limitations, such as limited degrees of freedom (DoF), and environmental constraints. PHTs can also be combined with haptic devices, enhancing haptic feedback perception, improving portability, reducing device size, weight, and power consumption, and minimizing hardware issues and maintenance needs [1].

These advantages and the recent increase in pseudo-haptics research, combined with new XR techniques and applications, motivated this new survey. While this paper presents a historical perspective and foundations, the survey focuses on recent years (Jan. of 2016 to Oct. of 2023). During the preparation of this systematic literature review (SLR), we found other surveys, such as one covering pseudo-haptics work published until 2020 [7]. Compared to other published surveys, our study includes recent articles (2021 to 2023) and diverse literature search methods, resulting in different articles not present in earlier surveys. The recent work from 2021 to 2023 represents 40% of the final selected items, focusing mainly on XR (79%) and mid-air interactions (68%). Additionally, we propose a taxonomy with additional categorization, further exploring multimodal interfaces. Our survey also emphasizes hybrid combinations and multimodal user interfaces, which are particularly relevant.

This survey begins by defining pseudo-haptics key concepts and providing a historical perspective and related work in Section II. Section III details the SLR method, based on the Preferred Reporting Items for Systematic Reviews and Meta-Analyses (PRISMA) methodology, due to its conceptual and practical advances for systematic review studies [8]. Here, we describe the key research questions, the literature search strategy, and the screening criteria adopted for the survey. Section IV discusses our literature review findings and proposes a taxonomy to address the proposed research questions. Each taxonomic category is described in a separate section. Section V describes pseudo-haptics visualization media, followed PHT Multimodalities in section VI. Section VII discusses the main application areas for PHTs. Section VIII addresses relevant challenges and limitations, suggesting potential future research areas, followed by the conclusion section.

## II. HISTORICAL PERSPECTIVE AND RELATED WORK

Objects and surface features influence sensory perception through their properties and appearance. You might recall the primary school question: "Which is heavier, 1 kg of iron or 1 kg of cotton?" This question explores cognitive factors influencing weight perception by comparing the appearance of materials with the same mass. In pseudo-haptics studies, a similar query can be made on an object's appearance and how it influences weight perception: "Which is heavier, a box filled with iron or the same box filled with cotton?" This example reflects how ecological psychology studies human perception and interaction with the environment. Ecological psychology explains the information transactions between living systems and their environments, and how they perceive significant situations to plan and execute actions [9]. It describes the historical perspective of ecological psychology, from Aristotle (c. 350 BCE) to Locke (1690), who posited that all knowledge comes from experience, allowing humans to gain knowledge through their senses. Perception evolves with Boring (1950), who suggested that a copy of the world exists between a person and the environment, enabling them to experience it [9]. This contact is mediated by a copy, informing behavior and influencing understanding of the world. When the copy is inconsistent, Helmholtz (19th century) suggested that ambiguities could be resolved through unconscious inference and the principle of maximum likelihood, creating an interpretable mental representation of the world for planning and executing behavior. The human mind examines incoming inconsistent stimulation, identifies cues and relevant rules, and unconsciously infers the world's state to explain that sensory stimulation [9].

Based on this mechanism theory, studies have long considered how faulty simulation interference can affect perception, by introducing illusions. Charpentier (1891) [10], published the first experimental work on the size-weight illusion (SWI) responsible for "disappointed expectations," where the speed of lift dominated perception. Sensory conflicts, and the dominance of visual stimuli over tactile, were empirically studied as early as 1964 [11]. They asked participants about the shape of a cube, with its visual appearance distorted through a lens, and how it differed from its tactile perception. In 2000, the term "pseudo-haptic feedback" was introduced to describe a haptic sensation, by combining user-applied forces in the virtual environment, with incongruent visual feedback from the simulation [6]. The discrepancy between applied forces and visual feedback, influenced by the user's prior knowledge, induces a manipulated perception of virtual objects and their properties. In pseudo-haptics, an artificial sensory conflict arises from combining multimodal cues (e.g., actual muscle tension with visually manipulated feedback). This conflict is resolved through visual dominance. The corrective process to maintain postural stability, known as compensatory postural adjustment (CPA), can induce an illusory sensation of perceived force, without a haptic device [1]. Multimodal sensory cues, such as object observation, skin deformation in response to surface pressure, or muscle and tendon tension, yield a combined mental representation of an acting force. These experiences align with ecological psychology studies, where affordance perception – the ability to perceive potential actions or uses of an object or environment – must be considered multimodal as everyday perception is multimodal [9]. Affordance perception involves visual and motor processing, and the integration of sensory information from multiple modalities. The nervous system does not perceive isolated properties. For example, distance is provided through eye muscles, body muscles kinesthesis, and proprioception, which continuously calibrate vision [12]. Although vision is often dominant, it is influenced by multi-modal perception, including the haptic component [9]. Besides being a multimodal perception mechanism, affordance perception is influenced by factors, including the individual's prior experience and knowledge, the context, and the individual's goals, action planning, and execution [9]. This survey explores how pseudo-haptic effects can influence the affordance features of an object, action, situation, or environmental perception, as individuals use their affordance perceptions to guide behavior and interact with their environment. This work's findings provide suggestions and recommendations for future research on influencing affordance perception, for example in XR environments.

## III. METHODOLOGY

This SLR aims to screen published literature on pseudo-haptics, applying a methodology based on the PRISMA framework [13], [14]. This methodology provides a standardized way to extract and synthesize information from existing studies related to research questions.

### A. RESEARCH QUESTIONS

As described in the previous section, the human mind examines incoming distorted stimulation, identifies cues, and the relevant rules, and unconsciously infers what must be happening in the world. As our research objectives, we hypothesize that by leveraging the mind's process of converting into something understandable, introducing PHT effects, that distort incoming environmental stimulation, can be applied in HCI design to simulate sensory perception effects [9]. Based on this hypothesis, we formulated the following research questions to delineate the problem being addressed:

**RQ1:** What are the main sense simulation effects used in PHTs? Rationale: By identifying the primary sense simulation effects currently used in PHTs (such as weight, texture, or resistance), we can better understand how these technologies manipulate user perception and which senses are most responsive to pseudo-haptic feedback.

**RQ2:** Which PHTs have been applied, and what are their main advantages? Rationale: This question aims to catalog the breadth of PHT applications, highlighting the strengths of these systems. Understanding their advantages can guide future PHT developments and help select the appropriate PHT for a given HCI design challenge.

**RQ3:** Which visualization media are being considered for visual pseudo-haptic perception? Rationale: Exploring the range of visual media used in PHTs (such as VR headsets, AR applications, or 3D displays) can reveal how different technologies affect the quality and efficacy of the pseudo-haptic experience.

**RQ4:** Which modalities are being combined? Rationale: Investigating the combination of sensory modalities, such as touch and vision or auditory and haptic feedback, is crucial for creating a holistic pseudo-haptic interface. This can lead to a deeper understanding of multi-modal integration and its impact on user experience.

**RQ5:** What are the main application use cases for PHTs? Rationale: By identifying and analyzing the primary use cases whether in gaming, medical simulations, or remote operations, we can evaluate the practical impact of PHTs and potential markets for these technologies. Understanding use cases is also instrumental in revealing the current limitations and areas for improvement in PHT design.

### B. SEARCH STRATEGY

The second step is to define a search strategy, and the adopted strategy, which includes determining the keywords and the semantics of the research. For this survey, in line with the research objectives, the search aims to be as broad as possible regarding pseudo-haptic description, applying a generic query: "pseudo-haptic". Next came the selection of digital libraries and publications to search for studies. Based on the available digital libraries, search engine capability, and publications (journals, conference proceedings, magazines), we considered the following databases: ACM Digital Library; IEEE Xplore; ScienceDirect; and SpringerLink.

### C. SEARCH AND SCREENING RESULTS

Fig. 1 shows the number of publications retrieved per year, using the keyword "pseudo-haptic," in all four databases considered. It shows a recent trend of increased research in this area, with a peak observed in 2023.

### D. SELECTION CRITERIA

The results obtained with the PRISMA methodology are summarized in the diagram flow presented in Fig. 2, which depicts the screening process and how we registered and analyzed the relevant information. To screen the initial 410 results, we defined the following selection criteria steps (exclusion, quality assessment, and inclusion), identifying primary references that would provide direct evidence for the research questions:
- Published in recent years, from Jan. 2016 to Oct. 2023 (#235 articles).

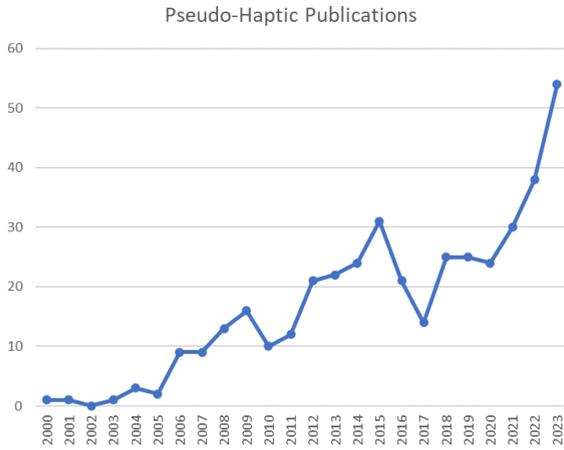

**FIGURE 1.** Number of publications per year, considering the keyword "pseudo-haptic" in the databases: ACM, IEEE, ScienceDirect, Springer.

- Written in English, excluding duplicated articles, reading the paper titles, and evaluating the abstracts (#66 articles). We found it most efficient to jointly read the paper titles and evaluate the abstracts in the same step, aiming to reduce article count by selecting articles with a primary focus on pseudo-haptics.

- Read the introduction and conclusions of each full article, ensuring a minimum of five pages (#47 articles). Articles with less than five pages were deemed too brief to deeply explore the topic. The introduction and conclusions were assessed to reduce the article count by selecting articles with a primary focus on pseudo-haptics. Survey articles were not considered in the taxonomy; instead, we included relevant original articles from the surveys, to avoid duplicate entries.

- Assessing full-text quality and eligibility (#42 articles).

The remaining articles were analyzed and read together. To assess the quality of the chosen articles, we used the following quantitative selection criteria, aiming to cover the research questions. Each paper received one point for each criterion effectively addressed. Only the articles scoring over 50% total points were retained, specifically those fulfilling at least three out of the five quality criteria:

*RT*- rationale (RQ1): Is there a rationale and discussion of the study's underlying assumptions? Does the study target a specific goal to be achieved?

*PT*- pseudo-haptic technique (RQ2, RQ3, RQ4): Are the PHTs clearly described? A clear description allows for better contextualization and assessment of outcomes.

*RC*- results conclusion (RQ2): Are the study results quantified and are the conclusions grounded on empirical research? While an empirical evaluation strengthens a given study's conclusions, we also considered papers featuring lessons-learned reports based on expert opinions, as they reflect end user perceptions.

*PA*- possible applications (RQ5): Does the research describe potential applications in industry, laboratories, and product usage or identify improvements? Can the findings be extrapolated beyond the original scope?

*LR*- limitations and research work (RQ2): Are the limitations of this study explicitly discussed? Does the discussion provide insights for possible follow-up research work or contribute to the broader discussion within the study area community?

The authors highlight that these full-text eligibility assessment criteria are connected to their own research proposal screening subjectivity and are not intended to represent a standalone article quality rating.

- Additional relevant articles reviewed: These include six new items (#48 articles). As we read the articles, any references to previously unselected papers, but relevant papers were added to the list (i.e., included as eligible).

The final selected articles for PHT analysis are summarized in Tables 1 and 2. Where for simplicity, we listed papers by their entry numbers (referring to each table's index column, e.g., 1, 2, 3, etc.).

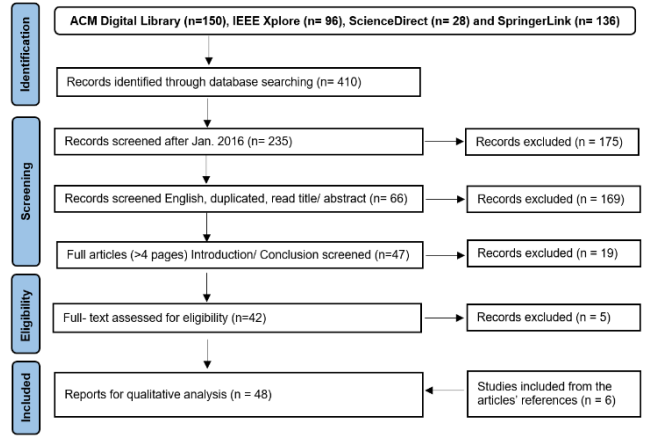

**FIGURE 2.** PRISMA diagram flow summary of the survey article screening process and obtained results.

## IV. PSEUDO-HAPTICS TAXONOMY

The proposed pseudo-haptics taxonomy themes consist of criteria to characterize how people perceive a sense simulation effect in haptic feedback. This is primarily done by observing the visual deformation technique, designed to distort a subject's illusion, directly mapping it with the "user action" input. Additionally, it considers whether different modalities were combined for a particular application. These elements can be considered part of a pseudo-haptics system architecture, illustrated in Fig. 3 block diagram proposal.

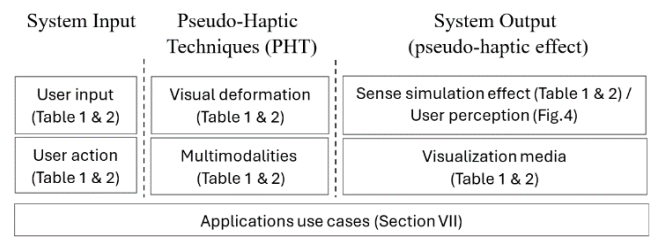

**FIGURE 3.** Pseudo-haptics system architecture diagram proposal.

TABLE I
PHT TAXONOMY: TACTILE FEEDBACK LITERATURE REVIEW.

| Table 1 index | Sense simulation effect ||||| Visual deformation ||| User action || User input ||||| Visualization |||| Multimodalities |||| Author, year |
|---|---|---|---|---|---|---|---|---|---|---|---|---|---|---|---|---|---|---|---|---|---|---|---|---|
| | Elasticity | Roughness | Stiffness | Friction | Stickiness | Displacement | Geometry | Appearance | Movement | Force | Mouse pad | Physical cue | Touchscreen | Muscle tension | Mid-air | Desktop display | Mid-air | HMD | Touchscreen | Haptic Cues | Actuated Displays | Vibratory/ squeeze | Acoustic/ sound | Human body | |
| 1 | 1 | | 1 | 1 | 1 | 1 | 1 | | 1 | | | | | | 1 | L | | | | | | | U | | Reynal et al. 2021 |
| 2 | | | 1 | | | | 1 | | | Ps | Ps | | | | | | | | 1 | | | | | | Yabe et al. 2017 |
| 3 | | | 1 | | | 1 | | | 1 | | | Vr | | | | 1 | | | | | 1 | | | | Onishi et al. 2021 |
| 4 | | | 1 | | | 1 | | | | 1 | Ps | | | | | | | 1 | | | | V | | | Mukashev et al. 2021 |
| 5 | | 1 | 1 | S | | 1 | | D | 1 | | | | 1 | | | Vp | | | | | | | | 1 | Sato et al. 2020 |
| 6 | | | 1 | | | 1 | | | | 1 | | Vr | | | | | | 1 | | | | Sq | | | Adilkhanov et al. 2020 |
| 7 | | | 1 | | | 1 | | | | 1 | | | 1 | | | 1 | | | | | | | | | Neupert et al. 2016 |
| 8 | | | 1 | | | 1 | | | 1 | | | | | | 1 | | | 1 | | | | V/Sq | | | Pezent et al. 2019 |
| 9 | | | | F | 1 | 1 | | | 1 | | | P | | | | 1 | | | 1 | | | | | | Ujitoko et al. 2019a |
| 10 | | | | K | | 1 | | | 1 | | | | 1 | | | | | | 1 | | | | | | Narumi et al. 2017 |
| 11 | 1 | | 1 | | | | | Pp | 1 | | | | | | | 1 | 1 | | | | | | | 1 | Kawabe 2020 |
| 12 | | | 1 | | | 1 | | | 1 | 1 | | 1 | | | | 1 | | | | | | | | | Fakhoury et al. 2017 |
| 13 | | | | K | | 1 | | | 1 | | | | 1 | | | | | | 1 | | | | | | Ban and Ujitoko 2018 |
| 14 | | | | K | | 1 | | D | 1 | | | | | | | G | | 1 | | | | | | 1 | Lécuyer 2017 |
| 15 | | Rf | | | | 1 | | | 1 | | | P | 1 | | | 1 | | | | | | V | | | Ujitoko et al. 2019b |
| 16 | | | Sp | | | 1 | | | | 1 | Sp | | | | | | | | | Sp | | | | | Chen et al. 2021 |
| 17 | | | 1 | | | 1 | | | | 1 | | | 1 | | | | | | 1 | | | V | | | Verona and Brum 2021 |
| 18 | | | | K | | 1 | | | 1 | | | | 1 | | | | | | 1 | | | | | | Hashimoto and Narumi 2018 |
| 19 | | 1 | 1 | F | | | | Vc | 1 | 1 | | P | | | | | | | 1 | 1 | | 1 | | | Rocchesso et al. 2016 |
| 20 | | Rm/Rf | 1 | F/S | 1 | 1 | 1 | | 1 | | | | 1 | | | | | | 1 | | | | | | Costes et al. 2019 |
| 21 | | | 1 | | | Cs | | D | 1 | | 1 | | 1 | | | 1 | | | | | | V | | | Li et al. 2016 |
| 22 | | | | 1 | | 1 | | | 1 | | | 1 | | | | | 1 | | | | 1 | V | | | Rietzler et al. 2018a |
| 23 | | Tp | | | | | | D | 1 | | | | | | F | | 1 | | | | | | 1 | | Tian et al. 2022 |
| 24 | | | 1 | | | | 1 | | | 1 | | | | | | 1 | | 1 | | | Pr | | | | Bouzbib et al. 2023 |
| 25 | | Tp | | | | | | D | | 1 | | | | | | 1 | | 1 | | | | V | | | Kim and Xiong 2022 |
| 26 | | T | | | | 1 | | F | 1 | 1 | | | | | | 1 | | 1 | | | | | 1 | Hb | Desnoyers-Stewart et. al. 2023 |
| 27 | | T | | | | | | | 1 | | | | | | 1 | | | 1 | | | | V | 1 | | Lee et al. 2023 |

| Rf- fine Roughness | F- Friction | D- Deforming | F- firework & color | Vr- VR controller | G- swipe Gestures | V- Vibratory | Sq- Squeeze |
| Rm-macro Roughness | S-Slippery | Vc- Visual Vibratory cursor | | P -pen | F- finger | Hb- heart beat ritem | |
| Sp- Spring | T- Touch | Cs- Cursor speed | Ps- Pressure sensor | | L- Levitated polystyrene | U- Ultrasound levitated polystyrene | |
| Tp- Typing | K - Kinetic friction | Pp- stretching & Poisson's ratio | | | Vp- Video projection | Pr- passive rigid prob | |

The inputs include the user's actions, as shown in Table 1 and 2. The PHT serves as the processing unit, designed to generate user perception effects from sense simulation. This is mapped directly to the proposed taxonomy in Fig. 4 and described in Table 1 and 2, alongside with the visualization media used. Those were based on the surveyed application use cases, as described in the dedicated section (addressing the RQ5).

Based on our findings, we propose a taxonomy of pseudo-haptic simulation effects according to user perception, as shown in Fig. 4. This taxonomy aims to provide a comprehensive visualization of the research question (RQ1). Pseudo-haptics simulation effects are usually mapped to an object's surface properties and geometry, but can also apply to a user's avatar.

For PHT tactile feedback on an object's or surface's properties, consider texture features such as stiffness, roughness, friction, stickiness, or elasticity. Kinesthetic feedback relates to the perception of weight and force. Attributes such as geometric properties, localization, orientation, angle, size, shape, or contact locations are relevant not only for objects but also for body ownership.

Pseudo-haptic feedback can also promote multisensory stimuli perceptions inherent to "object grabbing" and "body ownership", which involves composite either tactile or kinesthetic feedback perceptions. In Table 1, the PHT visual deformation techniques (second column) should be analyzed alongside the associated user input (third column).

PHT effects are achieved by combining visual deformation techniques with the person's input. In Table 1, the user carries out a physical action, such as a movement, an applied force, a pressing duration, or something else. These actions can use a mouse pad (e.g., 21), a physical cue (e.g., 8), a touchscreen (e.g., 7), muscle tension (e.g., 2), or mid-air (e.g., 1).

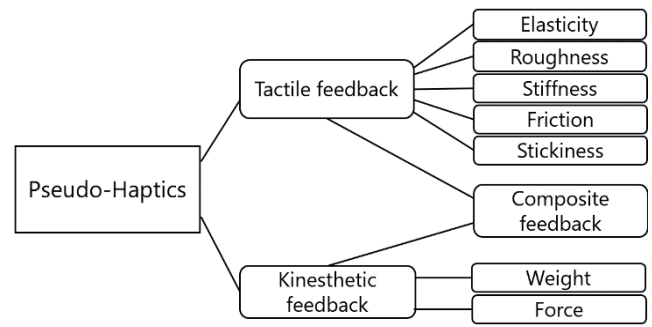

FIGURE 4. Proposed pseudo-haptics taxonomy, based on the user's perception.

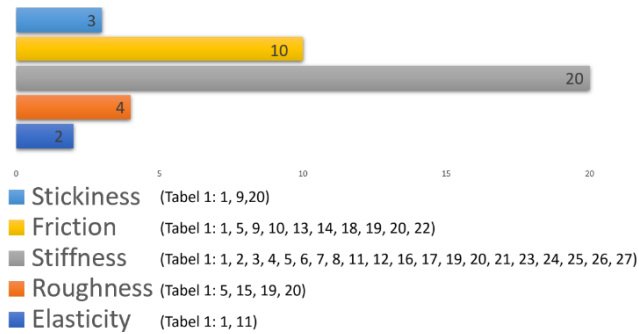

**FIGURE 5.** Tactile feedback survey occurrences distribution per pseudo-haptic effects (references to Table 1 index number).

Tables 1 and 2 show the PHT breakdown, where the pseudo-haptic visual illusion relies on deformations presented to the user as feedback related to their actions, such as:

- **Displacement.** Users visualize their actions with distorted displacement feedback on a display. This is used in 69% of the PHTs (Tables 1 and 2). For instance, a study concluded that the displacement technique is suitable for simulating haptic compliance sensations and has a just noticeable difference (JND) ratio to stimulus intensity similar to that obtained with haptic devices [15]. These promising results, also observed in other studies, likely explain the significant presence of PHT displacement distortion in the surveyed articles. The most common displacement illusion manipulates the control sensitivity in design, i.e. the control–display (C/D) ratio, which is the relationship between the user's input action, and the consequent visual change [16]. When the C/D ratio is 1.0, a direct relationship exists between the visualized artifact and the control device (e.g., they move at the same speed). When the C/D ratio exceeds 1.0, the artifact moves faster than the control device. When the C/D ratio is less than 1.0, it moves slower. Thus, a high C/D ratio means low control sensitivity, and vice versa. Altering the visual gain (i.e., the ratio between the user's input and the visual displacement) forces the user to resolve a visual-haptic conflict (e.g., CPA), making it possible to simulate sensation. This PHT is applied to translations, but occasionally to rotations (Table 2: 6, 18, 23). The visualized target could be an object (e.g., Table 1: 7), a screen pointer (e.g., Table 1: 21), or a body part (e.g., Table 1: 5).

- **Geometric deformation.** The appearance and size of objects or surfaces can influence perception. For instance, the size of an object can lead to overestimating its expected weight [17]. Several articles discuss geometric deformation (Table 1: 1, 2, 20, 24 and Table 2: 1, 4, 15, 18, 21, 29). Various techniques are described in [18], where haptic perceptions were represented by distorting motion or the shape geometry, as applied to a virtual cursor, while the person interacts with a touchscreen. Without any mechanical actuator, purely visual effects allowed people to perceive different haptic sensations (e.g., softness, stickiness, slipperiness, and roughness).

- **Appearance deformation.** Changing the visual appearance of an object, surface, or human body representation was present in 19% of the PHTs (Table 1 and 2). A person's understanding and knowledge of material properties can influence perception. The most direct PHT method is to modify surface texture appearance. A previous study [19], changed the perception of an object's stiffness by rendering it with two different meshes and textured static images. Participants perceived a more rigid rim rather than a softer tire. Changing an aluminum object's appearance caused it to be perceived as heavier, inducing a weight illusion. Appearance deformation can be also modulated by simply changing the color of an object, cursor, or the user's skin. Studies [20] and [21], explored manipulating visual feedback by superimposing a projected appearance onto the body or objects (e.g., by modifying perceived surface softness or by changing fingernail or finger color when touching an object). The amplified body variation visual feedback, makes the physical surface feel softer than with the normal white color. Also, in VR, choosing an avatar's stylized representation, rather than a realistic one, can also be used as a PHT. Exploring embodied metaphors [22], including abstract particle body representations, aura-like energy fields, fluid hands merging with other bodies, and dissolving bodies interaction patterns can simulate VR proximity, resistance, and contact.

A detailed description of our findings, and experimental results, is presented in the following sections grouped by each taxonomy category, addressing the first two research questions. Also, the taxonomy table allows for different reading flows serving as a tutorial. For instance, entry 14 in Table 1, demonstrates that visual appearance deformation can be used to induce friction sense simulation.

### A. PHT TACTILE FEEDBACK

The PHT tactile feedback category includes different sense simulation effects, as shown in Fig. 5. The most common perceptions were stiffness and friction, followed by roughness, elasticity, and stickiness.

#### 1) STIFFNESS

Stiffness relates to the perceived softness, hardness, and compliance of an object or a surface. [7]. It provides a sense of how much an object or surface can deform under force. Here compliance and deformation represent the extremes of a stiffness scale [23]. Stiffness was examined in 74% of the PHT tactile feedback studies (Table 1: 1 to 8, 11, 12, 16, 17, 19, 20, 21, 23 to 27).

A study [15] examined the visual impact of compliance in haptic discrimination. Their results based on Weber Fraction estimation suggest that a visual change of 19.6% from the reference compliance is needed to perceive a change in haptic compliance when using PHT. This falls within the 14–25% range identified in previous literature on compliance discrimination of deformable objects with haptic interfaces. This finding indicates that the pseudo-haptic visual displacement technique effectively simulates haptic compliance sensations, with JND ratio similar to that of haptic devices. These results, likely explain why stiffness was the most common PHT tactile feedback in the surveyed articles. Another PHT stiffness example is a physical pressure sensor, designed as a smartphone attachment to measure the user's squeezing force and its relationship to the induced visual pseudo-haptic softness distortion [24].

Recent papers focus on XR environments. For instance, in virtual typing [25] [26], a PHT virtual visual cue can reduce the uncorrected error rate without compromising typing speed. For VR touch, examples include VR avatar pseudo-haptics interpersonal social touch [22] and AR GUI interfaces [27]. PHT effectively simulates stiffness from 24 N/cm (e.g., a gummy bear to rigid objects) [28]. Studies comparing AR and VR environment studies [30] found that object stiffness perception depends on the XR environment. with similar objects reported as softer in AR than in VR. Thus when designing XR environments, consider that perceived stiffness may affected not only by the PHT but also by the XR environment in which the object is rendered.

### 2) ELASTICITY

Elasticity refers to an object's ability to return to its original shape after deformation [31]. A notable example of elasticity through appearance deformation with mid-air action is proposed in [32]. The first experience simulates stiffness perception, by pulling a virtual surface with a lateral hand movement in mid-air. The horizontal stretching of the display object varied in correlation with the distance between the user's hands. Manipulating the deformation-to-distance ratio, C/D ratio technique, showed that a smaller hand distance produced maximum deformation at larger ratios. The perceived effect of vertical compression (the Poisson effect) was evaluated, comparing rubber (high Poisson ratio) to cork (low Poisson's ratio). Experiments indicated that Poisson's ratio contributes to perceived softness, due to the brain's pre-knowledge of this relationship. These two PHT visual effects could be perceived as object elasticity, depending on mid-air hand gestures and how the object returns to its original shape after deformation. Therefore, they are categorized separately in this PHT taxonomy (Table 1: 1, 11).

### 3) FRICTION

Several articles study friction PHT tactile feedback (Table 1: 1, 5, 9, 10, 13, 14, 18, 19, 20, 22). In previous studies [6], the velocity of a cube in a virtual environment, was based on the user's input force and the intended friction coefficient. A similar technique [33], changes the displacement control ratio between finger movement on a touchscreen and background image scroll displacement. Users perceive strong resistance when the background image moves slower than their finger. Psychophysical results showed that with a C/D ratio of 0.34, users recognized the frictional feeling with a 75% probability for a single swipe, increasing to 79% for five repeated swipes. This suggests that perceived resistance feedback strengthens with repeated movements. This dynamic friction can also be induced with mid-air gestures [34], where an offset increase between the haptic cue tracker, and the virtual hand, depends on the simulated friction effect. Pseudo-haptic appearance deformation was used in a 3D carousel ring menu [35], where the GUI can repulse or attract attention to certain items during swipe gestures. Static friction pseudo-haptic methods allow users to perceive different material surfaces with various static coefficients. The stick-slip PHT [36] relates user input displacement while exploring virtual surfaces. During the stick phase, users perceived virtual contact points as stuck leading to a loss of felt agency. To address this, a virtual string visual cue extends from the stuck point, as the user moves the input device, maintaining the sense of agency. The length of this inconsistency is defined by the static coefficient; a higher coefficient makes the virtual pointer stickier. During the slip phase, the kinetic friction technique retards movement, based on Coulomb's stick-slip model. In a stick-to-slip transition, the virtual pointer should instantly translate to the user's touch point. Slip-to-stick transitions occur when the virtual pen's velocity is zero. Experiments showed that changing the static friction coefficient from 0.4 to 1.0, resulted in users perceiving pseudo-haptic static friction with over 90% probability and a 23% change in perceived friction intensity. Virtual spring visual cues were also used to enhance PHT friction effects on touchscreens [37].

### 4) STICKINESS

Stickiness was described in three of the surveyed PHT tactile feedback articles (Table 1: 1, 9, 20). Similarly to static friction, stickiness can be simulated using Coulomb's model [18]. PHT stickiness or slipperiness, as well as dryness or wetness, is simulated by displacement on a touchscreen, combined with cursor geometry deformation. Stickiness is represented by the deformation limit between the sticking and sliding phases.

### 5) ROUGHNESS

The roughness was described in four PHT tactile feedback studies: Table 1: 5, 15, 19, 20. Most roughness studies use the pseudo-haptic visual displacement technique, as seen in earlier research [38], [39]. In [40] the roughness was simulated with a stylus-based vibrotactile input device on a touchpad, with visual perturbation reflected in a cursor on the display. Users perceived the surface as rougher based on visual feedback parameters. This technique was also applied to touchscreens for simulating fine and macro roughness [18].

### 6) OTHERS

Although not present in this survey taxonomy (Tables 1 and 2), other PHT references are found in the literature, such as:

TABLE II
PHT TAXONOMY: WEIGHT, FORCE, OBJECT GRABBING, BODY OWNERSHIP.

| | Table 2 index | Visual deformation | | | User action | | User input | | | | | Visualization | | | | Multimodalities | | | | | Author, year |
|---|---|---|---|---|---|---|---|---|---|---|---|---|---|---|---|---|---|---|---|---|---|
| | | Displacement | Geometry | Appearance | Movement | Force | Mouse pad | Physical cue | Touchscreen | Muscle tension | Mid-air | Desktop display | Mid-air | HMD | Touchscreen | Haptic Cues | Actuated Displays | Vibratory/ squeeze | Acoustic/ sound | Human body | |
| Weight | 1 | 1 | 1 | | 1 | | | | | | 1 | B | | | | | | | U | | Reynal et al. 2021 |
| | 2 | I | | | 1 | 1 | | | | 1 | 1 | | 1 | | | | | | | M | Rietzler et al. 2019 |
| | 3 | 1 | | | D | | | V | | | | 1 | | | | | 1 | | | | Onishi et al. 2021 |
| | 4 | Rt | 1 | | Rt | 1 | | | | | 1 | | 1 | | | | | | | M | Lee et al. 2019 |
| | 5 | V | | | 1 | | | G | | | 1 | | 1 | | | W | | | | | Hirao et al. 2020 |
| | 6 | Rt | | | Rt | | | 1 | | | | | 1 | | | | | | | | Yu and Bowman 2020 |
| | 7 | 1 | | | 1 | | | | 1 | | | | | | 1 | | | | | | Ban and Ujitoko 2018 |
| | 8 | 1 | | 1 | 1 | | | S | | | 1 | 1 | | | | 1 | | | | | Lécuyer 2017 |
| | 9 | 1 | | | 1 | | | 1 | | | | | | 1 | | Wg | | | | | Samad et al 2019 |
| | 10 | I | | | 1 | | | V | | | 1 | | | 1 | | | | | | | Rietzler et al. 2018b |
| | 11 | 1 | | | 1 | | | | | | 1 | | | 1 | | W | | | | E | Kim et al. 2022 |
| | 12 | 1 | | | 1 | | | 1 | | | | | | 1 | | W | | | | | Feick et al. 2022 |
| | 13 | 1 | | | 1 | | | | | | 1 | | | 1 | | Vt | | | | | Hirao et al. 2023 |
| | 14 | 1 | | | 1 | | | | | | 1 | | | 1 | | | | | | T | Stellmacher et al. 2023 |

I- Inertia  D- open door  G- game control stick  B- levitated polystyrene  W- physical weight  Vt- VR controller trigger resistance
V- velocity  S- passive stick  U- Ultrasound Levitated Polystyrene  E- EMS  M- Muscle tension sensor
Rt- Rotation  V- VR controller  Wg- weight cube reference (optically tracked motion gloves)  T- Tendon Vibration

| | | | | | | | | | | | | | | | | | | | | | |
|---|---|---|---|---|---|---|---|---|---|---|---|---|---|---|---|---|---|---|---|---|---|
| Force | 15 | D | S | | 1 | 1 | | Rb | | | | | M | | | | | V | | | Osato and Koizumi 2018 |
| | 16 | In | | | 1 | 1 | | | | 1 | 1 | | 1 | | | | | | | Mt | Rietzler et al., 2019 |
| | 17 | 1 | | | 1 | | | V | | | | 1 | | | | | 1 | | | | Onishi et al. 2021 |
| | 18 | 1 | 1 | | Rt | 1 | | | | | 1 | | 1 | | | | | | | Mt | Lee et al. 2019 |
| | 19 | U | | | 1 | | | 1 | | | | | 1 | | | | | | | | Kang et al. 2019 |
| | 20 | R | | | 1 | | 1 | | | | 1 | 1 | | | | | | | | | Li et al. 2019 |
| | 21 | V | 1 | | 1 | | | | 1 | | | | | 1 | | | | | | | Costes et al. 2019 |
| | 22 | W | | | Vg | | | | | | W | | 1 | | | E | | | | | Hirao et al. 2022 |
| | 23 | Rt | | | | T | | | | | 1 | | 1 | | | K | | | | | Feick et al. 2023 |

D- force direction  W- Walk in place  Rb- small robot  M- mid-air imaging  Mt- muscle tension sensor
In- inertia  Rt- rotation  T- torque  V- VR controller  V- vibrotactile small robot
U- underwater drag force  Vg- velocity gain  E- elastic strips
R- upper-limb force rehabilitation  K- nnobs
V- viscosity  S- shrinking for adding force in the movement reference

| | | | | | | | | | | | | | | | | | | | | | |
|---|---|---|---|---|---|---|---|---|---|---|---|---|---|---|---|---|---|---|---|---|---|
| Object grabbing | 24 | 1 | | | | | | | | | 1 | | | 1 | | T | | | | | Baldi et al. 2021 |
| | 25 | | L | | | G | | | | | | | S | 1 | | | | 1 | 1 | | Haruna et al. 2021 |
| | 26 | 1 | | 1 | | Tg | | S | | | | | H | 1 | | Ge | | | | | Lécuyer 2017 |
| | 27 | 1 | | | 1 | | | C | | | 1 | | | 1 | | | | | | 1 | Azmandian et al. 2016 |
| | 28 | 1 | | | 1 | | | 1 | | | | | | 1 | | 1 | | | | | Feick et al. 2022 |
| | 29 | | 1 | | | 1 | | | | | 1 | | | 1 | | Pr | | | | | Bouzbib et al. 2023 |

L- robot finger visual light  S- passive stick  S- mechanical sensors/ hands wearable  T - passive fingers thimbles
G- Grasping  C- single physical cube  H- Hand and finger  Ge- passive grip elastic device
Tg- touch and gripping force  Pr- passive rigid prob

| | | | | | | | | | | | | | | | | | | | | | |
|---|---|---|---|---|---|---|---|---|---|---|---|---|---|---|---|---|---|---|---|---|---|
| Body ownership | 30 | H | | D | 1 | | | | | | | F | S | | | | | | | 1 | Kanamori et al. 2018 |
| | 31 | H | | D | 1 | | | | 1 | | | | V | | | | | | | 1 | Sato et al. 2020 |
| | 32 | 1 | | 1 | | T | | S | | | | H/F | | 1 | | G | | | | 1 | Lécuyer 2017 |
| | 33 | 1 | | | 1 | | | C | | | | A | | 1 | | | | | | 1 | Azmandian et al. 2016 |
| | 34 | G | | | 1 | | | | | | | A | | 1 | | | | | | | Misztal et al. 2022 |
| | 35 | Se | | | 1 | | | | | | | Se | | 1 | | | | | | | Li et al. 2022 |

H- virtual hand  D- deforming  S- passive stick  F- finger  S-stereoscopic projection  G- passive grip elastic device
Se- shoulder & elbow  C- single cube  H- hand  V- video projection
G- ghost effect  T- touch and gripping force  A- arm

- **Temperature.** Could observing the "Inverno" painting by Maria Vieira da Silva make one feel warmer on a cold winter day? Study [41] supports the theory of color-associated thermal perception, a long-standing tradition among artists. Experiments showed that red is perceived as warm/hot (64%) and blue as cool/cold (59%), illustrating color-specific psychological responses. Integrating the impacts of color into indoor environments can psychologically manipulate perceived temperature, offering insights for architects and indoor designers to achieve low-energy designs, and incorporate PHT.

Another proposed temperature PHT, uses a Peltier device to create a noticeable temperature change synchronized with arm movement [42].

- **Dynamic luminance.** This feature is crucial in deformation lamp techniques. Dynamic luminance projection makes static objects appear dynamic, activating motion and shape perceptions in the visual observation system. For example, making a static fish look dynamic while preserving the color and texture of the original object [43].

### B. PHT KINESTHETIC FEEDBACK

The PHT kinesthetic feedback, as described in the proposed taxonomy in Fig. 4, comprises weight and force.

#### 1) WEIGHT

The PHT weight articles are presented in Table 2 (1 to 14). Psychophysical experiments suggest that users could effectively "feel" weight through PHT, especially when comparing objects as lighter or heavier. The perceived weight of an object depends on several factors [2]:

- Touch Contact Area: Weight discrimination is more effective when lifting an object rather than just holding it. Lifting involves, two opposing forces: gravity pulling the object down and muscle force lifting it.
- Force Perception Distortion: Factors like muscle fatigue, tactile sensitivity, and gripping method, can distort force perception.
- Visual Characteristics: Size, shape, temperature, color, and brightness can affect weight perception. A slippery object is also perceived to be heavier with the additional grip force required.
- Mass Distribution: Represented by the object's rotational inertia and resistance to rotational acceleration.
- Sensorimotor Memory bias: Influenced by previous lifting experiences and implicit knowledge.

From the surveyed articles, different weight simulation techniques were found.

**Control/Display Ratio and Velocity**. Object behavior may influence the perception of heaviness more than their static appearance [2]. Most reviewed articles used C/D ratio control to distort visual displacement instead of user input. Smaller displacement, makes the virtual object feel heavier. This can be manipulated through various C/D ratio implementations, as lighter objects are easier to move. Identified techniques include:

*-Discriminated perceived weight*: In [44], participant's motions were captured by gloves with an HMD setup. They lifted a passive haptic (tangible) cue object with a reference mass of 185g (common for VR controllers). Users perceived weight changes through visual hand position manipulation to induce an illusory weight perception. Experiment results showed that ±5 g variation could be perceived using pseudo-haptic visual feedback, less than 5% of the reference weight. indicating a fine granularity in weight perception. This effect corresponded to a hand displacement of 5–10 cm, crucial to preserve the user's sense of agency and presence. A recent study [45], showed that the sensitivity to weight differences decreased as the C/D ratio increased. Subjects were more sensitive to small changes in C/D ratio when the standard object had a lower C/D ratio, relevant for XR designs involving multiple weight experiences.

- *Object inertia*: Inerta related to its weight, can be perceived using the C/D ratio control technique [46], [47]. The virtual objects' inertial properties are designed by displacing the real positional tracking of the user's hands or VR controllers when maneuvering heavier virtual objects. This approach requires more time to accelerate and decelerate movement, due to inertia. The study introduced a "time deformation" technique, enhancing the realistic perception of inertia. Experiments showed that distorted tracking offsets induced users to raise their arms higher as the tracking offset increased, creating an immersive illusion of a 'bowling ball' weight in a VR game.

*-Object mass distribution and rotation movements*: The C/D ratio control can modify object mass distribution [2], having a larger effect on perceived heaviness. When a rotational gain effect is applied to virtual objects, faster rotation makes the object feel lighter. Another method involves scaling rotational motion C/D ratios in a VR environment, reducing the rotational angle for heavier objects [48]. Psychophysical experiments showed that users perceived the object as heavier by more than 80% when using eight different C/D ratio levels.

- *XR controller-based versus motion tracking-based*: Weight perception can emerge from velocity visualization, through virtual object motion or the user's hand movement. This concept was explored by comparing motion-based and controller-based methods, while participants drew or lifted a virtual object [49]. The study used seven delay parameters to simulate different virtual weights in a VR environment. Participants dragged a virtual object, by either extending their hand to grab the box handle and pulling it towards their body (motion-based manipulation), or by tilting controllers to manipulate the virtual hand movement without physical movement (controller-based manipulation). In this method, the object moves slower when close to the target, and faster when farther away. Researchers adopted a hand range of 0 to 2 m/s, tuned that appear natural. The perception of virtual weight with pseudo-haptic methods was successfully represented by adjusting the object's velocity for both controller-based and motion-based manipulations. The results support the hypothesis that sensory information (e.g., vision, audition, touch, and kinesthetic sense) is integrated and combined, to produce an optimal estimate in the brain. Due to visual cue dominance, both approaches successfully presented pseudo-weight sensations in VR, even when the user's body was not moving (e.g. using only a game controller).

*-Mid-air actions*: Research [49] highlights flexibility for GUI designers in various XR applications, suggesting that PHT weight perception can be effectively induced by C/D ratio and velocity, without requiring physical arm or hand movement. Instead, the virtual representation of those movements suffices.

The survey also indicates a significant trend of using pseudo-haptics in mid-air actions in XR by mapping effects directly to physical body movements [2]. This presents a research opportunity to explore weight perception through mid-air technology known for its flexibility and adaptability. Recent studies [50] leverage lifting force by tracking hand position and rotation, using a force-arrow indicator to guide the person on increasing or decreasing the speed. This method induces slower hand movements for heavier objects or faster for lighter ones, achieving a 17% increase in weight distinction accuracy without requiring an HMD.

Another study [51], proposed pseudo-haptic visual speed illusion feedback, where users lift virtual objects using mid-air gestures. This method provides a pseudo-heaviness effect by making finger gestures to pull up concentric grating stimuli on the display. As visual feedback magnitude decreases, the sensation of heaviness increases.

In a related work [35], participants used a passive haptic (tangible) stick and observed their avatars in a VR environment. Real-time mapped mid-air gestures to distorted avatar animation. This modified motion amplification, or reduction, by distorting temporal motion, or changing the avatar's posture. The primary PHT approaches for creating the illusion of weight include movements, C/D ratio, and velocity technique, though we describe below other less-used methods.

**Shape**. Some research investigated how changes in shape without altering volume, affect weight perception. Objects were formed into various two-dimensional figures, such as circles, squares, and irregular polygons. Results [52] suggested that more compact shapes were perceived as heavier than less compact ones (e.g., spheres were perceived as heavier than tetrahedrons and cubes; cubes were heavier than tetrahedrons).

That study also suggested that shape affects perceived weight similar to size. That implies that the influence of shape on perceived weight may depend on perceived size. Generally, larger objects are assumed to be heavier than smaller ones. The SWI illusion notes that the perceived weight of an object is influenced not only by its actual weight but also by its size. When lifting two objects with identical mass but different volumes, the smaller object is usually perceived to be heavier. The SWI can reduce perceived heaviness by 50% when increasing the size without changing the mass.

**Size**. Object size significantly influences perceived weight. Studies found that variations in an object's physical size resulted in more noticeable differences in perceived weight than changes in its apparent material [53]. In one experiment, participants lifted virtual dumbbells of different sizes, affecting both weight perception and lifting speed [50]. Another evaluation considers how virtual object size influences the distortion value required for weight perception [54]. The conclusion was that participants more readily perceived weight in smaller virtual cubes. As the cube size increases, visual pseudo-haptic distortion must also increase for participants to perceive additional weight. Thus, the C/D ratio needs adjustment not only for the intended weight simulation but also for the target object's size.

**Audio**. Other studies examined how audio effects influence weight perception, such as whether a full-body avatar's perceived weight in VR can be affected by different footstep audio effects. Avatars were judged as being heavier when audio filters amplified lower center frequencies. Other audio features, such as a louder and faster heartbeat were used to represent increased force magnitude [2].

**Color and Brightness.** These qualities can influence sensory perception. Light-colored objects are often perceived as slightly heavier than darker ones. However, the influence of an object's color on its weight is inconsistent and relatively slight compared to other illusions. Brighter objects are generally seen as lighter, but the brightness–weight illusion is less robust and generalizable than the size–weight, material–weight, and shape–weight illusions [2].

**Temperature**. Temperature also appears to influence object weight perception. The temperature–weight illusion causes cold objects to be perceived as heavier [55]. Psychophysical experiments on the effect of temperature on the perception of heaviness, grasp, and lift forces, were conducted using cold (18 °C), thermal-neutral (32 °C), and warm (41 °C) objects, with two distinct masses (light: 350 g and heavy: 700 g). Cold objects felt 20% heavier than thermal-neutral ones. Additionally, grip force increased by 10% when cold objects were lifted compared to thermal-neutral ones. These findings suggest that cooling an object not only increases its perceived weight but also influences applied force during grasping and lifting. This approach opens up research opportunities for multimodal techniques.

**Physical Haptic Cue.** Most studies induced perceived weight through distorted displacement while participants held a physical (tangible) cue, with a weight reference point [49], [44]. However, some research has shown that the same effect can be achieved without physical weight cues, as shown with mid-air interfaces [23].

### 2) FORCE

Humans sense weight by integrating proprioceptive, cutaneous force feedback, and tactile information. For instance, when lifting an object, weight perception is derived from these combined forces [2]. Typically, these forces are simulated using ground-based kinesthetic force feedback actuators or wearable haptic devices, which provide soft kinesthetic and tactile sensations, often in the form of an exoskeleton. In addition to haptic force-feedback devices, visual pseudo-haptics can also simulate force perception.

The surveyed articles on PHT force are listed in Table 2 (15 to 23). For instance, a perceived torque force (ranging from 6.683 to 12.684 Nm) was achieved through VR knob rotational C/D ratio manipulations [29]. Another example is creating the sensation of virtual wind on a user's hand by visualizing an object's velocity and its perceived mass and weight [1]. The PHT effect was induced by distorting the hand's position on the visual display. When a user tries to resist a simulated force by maintaining their hand position within the force field, the visual display shows the hand involuntarily moving with the virtual flow. This causes the user to stabilize their hand position, often resulting in compensatory movements in the opposite direction. The simulated displacement represents the virtual force, and the corresponding real-hand effort creates the illusion of an applied force. Another method simulated virtual kinesthetic forces by combining muscle tension input with pseudo-haptic feedback [47]. Heavier objects require more time to accelerate and decelerate due to inertia.

A force-arrow method indicates whether the user needs to increase the lifting force [50]. In upper-limb rehabilitation, virtual motion assistance guides movement along a path and creates the illusion of a reacting force [56]. The application guides movement along a path. By manipulating the ratio between limb movement and virtual cursor speed, users feel objects as lighter or heavier based on cursor movement. Two experiments used a computer mouse and a mid-air interface via a depth sensor. The closer the cursor is to the exit path of a virtual maze, the faster it moves. Virtual resistance increases when the arm strays from the path's center, making it harder to keep the cursor centered due to arm weakness or hand trembling. This displacement is perceived as additional force, complicating cursor control.

The underwater sensation of viscosity related to drag force can be simulated using PHT [3], applied to human limbs during swimming. The goal was to determine how much drag, indicated by visual displacement distortion, could be applied to provide a natural underwater sensation and enhance immersion. They estimated a drag-force multiplier value from a transfer function experiment. The x-axis represented the drag force multiplier, and the y-axis the probability that a drag force was judged as "natural." Results approximated a Gaussian distribution, with natural multipliers between [2.5, 5.6] and a standard deviation of 1.56. They took the mean (4.05) as the most natural pseudo-haptic underwater drag force multiplier, which is much higher than in an on-land situation. Fluid viscosity is similar to underwater drag force, where a density force distorts cursor movement [18]. The viscosity force is opposed and proportional to finger movement over a touchscreen. Fluid viscosity can be simulated by the viscous resistance model (1), where r is the cursor size, η is the fluid viscosity, and V is the uniform stream velocity [7].

$$F = 6\pi r \eta V \qquad (1)$$

Pseudo-haptics can also simulate force effects in push and pull interactions. For example, in a game, the visual representation of a computer-generated character (CGC) movement, was combined with a small robot to control the CGC indirectly [57]. The authors experimented with the pushing and pulling movements of the CGC. A movement of the CGC was performed by pushing the small robot while shrinking, where the shrinking representation acted as a proxy for the applied force. Another movement was performed by hitting the robot while rotating. Participants were asked about the direction of the force. Results showed that pseudo-haptic feedback applied to the CGC's visual representation effectively altered the perceived force direction. A similar type of push and bump simulation is proposed using the "BouncyScreen" approach [58]. Here, a person manipulates a virtual object with a VR controller, synchronized and combined with the visual C/D method. Their findings suggest that induced force feedback by physical screen movement was almost as effective as the vision-based pseudo-haptic approach. Additionally, "BouncyScreen" provided different force representations depending on interaction types (e.g., pushing and bumping), significantly enhancing the reality of the interaction and sense of presence.

### C. COMPOSITE FEEDBACK

This taxonomic category considers feedback promoting multisensory stimuli perception during "object grabbing" and "body ownership" actions, involving composite tactile and kinesthetic feedback (see Fig. 4).

#### 1) OBJECT GRABBING

This category addresses the action of grabbing or grasping an object. Table 2 (24 to 29) presents the PHT Object Grabbing articles surveyed.

Two sensations convey weight when humans grab and lift objects [2]. First, skin pressure and grip mechanoreceptors provide force sensation. Second, proprioception gives information from muscle and tendon strain. However, these mechanisms are absent when manipulating remote objects, e.g., in XR. The proposed self-contact method of pseudo-haptic kinesthetic feedback aims to overcome the lack of tactile information in virtual object-grabbing [59]. This method explores the C/D ratio difference between human hand velocity and virtual hand velocity until the user makes contact between their fingers during grasping tasks. The study concluded that kinesthetic feedback increases the realism of grasping, and physical contact between fingers helps users maintain the grasp, making the task easier. Another study on 3D object manipulation combined a handheld passive elastic input device, allowing users to touch and apply grab force to manipulate 3D objects, together with pseudo-haptic feedback simulating different grasping effort levels [35].

In the study on how grasping types affect PHT in VR [60], users choose grasp types based on task requirements, such as power grasps for heavier/ larger objects and fine precision grasps for smaller ones, and object characteristics (shape, size, and mass). The results indicate that for handheld objects (<500 g), variations in object shape, grasping type, or mass do not significantly affect perceived disparity.

In a remote humanoid robot maneuvering scenario, the robot arm grasps an object controlled by the user's hand through mechanical detection sensors or a hand-wearable measurement device [61]. When the pressure sensor on the robot hand's gripper exceeds a threshold, the operator receives a visual light feedback indicator on the robot's fingertips. Haruna et al.'s [61] results suggest this method is effective in minimizing the exercised grasping force and reducing the processing load on the operator's brain.

### 2) BODY OWNERSHIP

The PHT body ownership articles, detailed in Table 2 (30 to 35), investigate the sensation of body ownership through synchronized visual and tactile stimuli. This illusion makes a person feel that a non-bodily object is part of their own body, as demonstrated by the rubber hand illusion (RHI) [62]. Another study expanded on the RHI by creating a six-digit hand illusion in VR, which participants accepted as part of their body to some extent when wearing an HMD [35]. These body ownership illusions, such as RHI, delve into the brain's ability to integrate multisensory information and form a coherent sense of self-awareness. The RHI, for example, uses visuo-tactile stimulation to make subjects perceive a rubber hand as their own by simultaneously stroking it while the real hand is hidden. Similarly, observing another person's actions can activate the observer's motor circuits, mapping observed actions onto their motor representations. This suggests that the brain represents others' actions as its own. However, factors like different viewpoints, morphological features, and kinesthetic experiences make exploring this phenomenon in VR challenging [62].

The most common PHT presents a distorted visual displacement related to the natural user action movement and position. This visual stimuli displacement can be applied not only to virtual object representations, but also to the human body part depictions, including fingers, hands, and arms, providing people with specific target haptic properties when interacting with the environment. VR can provide an immersive experience of body ownership illusion as affecting the user's cognition, perception, emotion, and behavior, by modifying virtual motion displacement, between the key joint positions of the user's physical body and the avatar's virtual body [63].

A similar PHT illusion was also used to simulate a virtual impairments function, as an applicable substitution for an arm physical splint of an age simulation suit. The arm VR movement impairment was obtained by using a visual tracking offset and a transparency effect, the "ghost effect". Results showed that PHT can effectively replace and provide a comparable experience as with a physical haptics age simulation suit, without negatively influencing general self-presence, and simulator sickness, while offering more flexibility and adaptability with users [64].

Another study [65] explored skin deformation as a method to provide effective pseudo-haptic feedback for force, stiffness, and friction, substituting kinesthetic force feedback in weight perception. The authors concluded that while humans can perceive weight through kinesthetic feedback and skin stretch cues in a virtual environment, they are less sensitive to skin deformation. Their experiment showed that the Weber fraction (the ratio of the JND to the stimulus intensity) was 11% for kinesthetic feedback and 35% for skin deformation feedback.

Further research [66] suggests that non-corporeal events can be perceived as body parts. To create a convincing illusion of body ownership, the design must systematically synchronize and map changes to the user's actions. These approaches will be further analyzed in this survey.

In [67], the body ownership illusion was explored while manipulating virtual objects. A haptic retargeting experiment demonstrated that a single passive physical cube prop can appear as multiple virtual objects. This technique dynamically aligns physical and virtual objects, leveraging the dominance of vision when senses conflict. For example, users believe they are moving their arm to reach different objects, but their arm consistently reaches the same physical cube. This effect combines two techniques: world manipulation, where the virtual world moves to align with the passive haptic prop, and body manipulation, where the virtual representation of the user's body adjusts to align with the prop. Results indicated that all haptic retargeting techniques enhance the sense of presence compared to typical wand-based 3D control of virtual objects.

Beyond HMD devices, body-ownership manipulation was also achieved using stereoscopic projection mapping [68]. By projecting the image of a touching finger onto a real object, spatial AR visually deformed the touched surface, influencing body ownership. The touching finger region, extracted from the captured image, was deformed and projected onto the displayed image, creating a geometrically accurate representation for both stationary and head-tracked observers. Psychophysical studies concluded that the perceived virtual shape lay between the physical and projected shapes, with the curvature radius effect being more substantial than the height effect.

Additionally, the human body augmentation method was applied using a touchscreen [69]. A mid-air projected virtual hand approached and overlaid real or virtual distant objects as if touching them, aiming to provide touch texture sensations of slipperiness, roughness, and softness. Experimental results indicated that users could perceive tactile sensations at five intensity levels solely from visual information, without any haptic devices. These studies suggest that mid-air projection of the virtual human body preserves the user's sense of ownership, and synchronized pseudo-haptic simulation reinforces body ownership.

## V. PSEUDO-HAPTICS VISUALIZATION

This section addresses RQ3, concerning the visualization media considered for visual pseudo-haptic perception. Since PHT simulates haptic feedback using vision, assessing the visualization mediums used is crucial.

The surveyed articles, summarized in Tables 1 and 2, confirm PHT usage across various visualization media. Recent studies have extensively explored sophisticated apparatuses for AR/VR applications and HMD devices (54%), followed by desktop displays (19%), touchscreens (17%), and mid-air setups (10%).

### A. HMD

Although HMD experiences are primarily visual, they are often combined with other modalities such as passive haptic cues, active vibratory or squeeze devices, and sound. The interaction with the human body is the most thoroughly explored aspect. These multimodal inputs, along with the user's physical sensory-motor actions, create a spatio-temporal sensory conflict between the virtual motions, distorted synchronously. Considering PHTs can make XR environments more immersive. Typically, user interaction in a VR system requires an input device providing precise input and haptic feedback. However, these devices do not convey a real-world interaction sense. Ideally, an immersive VR environment should allow user interaction as naturally as body movement. Some surveyed articles describe applications combining HMDs with mid-air interfaces, eliminating the need for external props as input devices or to receive PHT kinesthetic feedback (Table 1: 8, 14, 23 to 27). This immersion can also leverage pseudo-haptics to create sensations of weight (Table 2: 2, 4, 5, 10, 11, 13, 14), force (Table 2: 16, 18, 22, 23), object grabbing (Table 2: 24 to 27, 29), and body ownership (Table 2: 32 to 35).

By creating a sensory illusion of an alternate reality, such as an avatar located in a different environment or location or interacting with remote or virtual objects, XR provides a sense of presence and immersion. However, HMDs pose a significant challenge due to the lack of haptic feedback, especially during mid-air interactions without a controller. Pseudo-haptics were applied to enhance GUI capability, task workload, motion sickness, and more immersive perception through mid-air finger-based menu interactions [70]. In another study, making physical contact between fingers during grasping tasks, increased realism, making it easier to maintain grasp on an object, thereby overcoming the lack of proper tactile feedback during virtual object-grabbing tasks [59]. Additionally, pseudo-haptics were explored in different contexts and scenarios, improving interactions in 3D and VR environment applications [35].

### B. TOUCHSCREEN

Touchscreens were used for PHT kinesthetic feedback (Table 1: 2, 9, 10, 13, 17 to 21), weight (Table 2: 7), and force (Table 2: 21). Nowadays touchscreen surfaces are widely used in devices such as tablets, some laptops, or smartphones. These touchscreens can act as a combined device by generating haptic effects on a physical display. These effects include vibrotactile displays or dynamic surfaces (variable friction devices or shape-changing surfaces) that stimulate biological sensing effects in a person's hands. For instance, the "Yubi-Toko" touchscreen system allows users to walk in a snowy virtual scene and feel the difficulty of moving forward due to the snow [98]. Based on the pseudo-haptic displacement technique, the visual ground scrolls according to the C/D ratio, combined with the visualization appearance manipulation, where the footprint is shown as dirty snow in dark contrast. In another study, induced variable friction can be based on device lateral forces (electrostatic implements or acoustic waves stimulating the fingertips) that can simulate material properties via the display. Additionally, shape-changing surfaces (e.g., pin-array tactile dynamic Braille displays, piezoelectric contractors) can create compression and traction effects [71]. For example, the E-Vita touchscreen provides friction haptic feedback directly on the device display [72]. This technology allows E-vita to enhance teleoperation situation awareness when an unmanned ground vehicle (UGV) loses traction. Due to the drawback of touch panels providing poor haptic feedback, they are often complemented with added haptic functionality, such as vibratory feedback (on the entire display or specific areas), mechanical feedback (e.g., a pen), or dynamic surfaces that mechanically affect the perception of texture on the touchscreen. However, like other haptic devices, touch panels face similar constraints and disadvantages, including cost, accuracy, power consumption, size, complexity, and limited haptic sensations. PHTs can significantly improve the HCI experience, as described in previous sections. Based on the analyzed articles and their experimental results, various sensory experiences can be simulated using PHT, enhancing the HCI experience on touchscreens. The studies primarily utilized visual pseudo-haptics in isolation. However, some experiments also combined it with auditory feedback [73].

### C. MID-AIR VISUALIZATION

Some surveyed articles did not employ a visualization media device. Instead, the pseudo-haptics were directly rendered in mid-air (Table 1: 1, 5 and Table 2: 1, 15, 30, 31). Mid-air hand gesture-based interactions in a virtual environment pose challenges, including missing haptic feedback, often making complex dynamic gestures frustrating (e.g., grabbing or squeezing an object). Pseudo-haptic visual cues can help overcome this limitation [74].

Comparing visual pseudo-haptics for dynamic squeeze or grab gestures in immersive VR, with and without pseudo-haptics, Ahmed et al. [74] concluded that pseudo-haptics enhance usability and user experience. The quantitative user study showed reduced time to perform gestures and fewer errors. The study of hand gestures in mid-air use cases is increasingly relevant for creating more natural immersive XR environments, and it is also a subject of our survey. Mid-air interfaces are not limited to VR. Interesting research has been published on direct mid-air interactions, such as [23], which describes an acoustic levitation display in mid-air, similar to a hologram. Consequently, users can interact with contactless mid-air gestures. Another approach involves mid-air imaging technology formed by incident light placed next to real-life objects to create CGCs that appear to pop off the screen [57]. One of the main advantages of mid-air visualization media is its potential to enhance users' sense of reality in virtual environments.

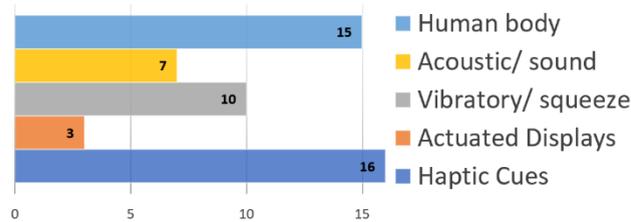

**FIGURE 6.** Multimodal occurrence distribution (Table 1 and 2).

## VI. PSEUDO-HAPTICS MULTIMODALITY

Addressing RQ4, pseudo-haptic stimulation methods can use visual PHT, but can also be designed with various other modalities. The distribution of modalities in the surveyed articles is shown in Fig. 6. It indicates that haptic active devices with actuators (vibration/squeeze), the human body (e.g., muscle), and passive haptic cues (e.g., physical weights) were the most commonly used modalities. Additionally, less commonly used modalities included acoustic/sound stimuli to present sensory stimulation directly to users and actuated physical display movements (actuated displays).

Modulating a specific haptic property with physical stimuli can result in a combined technique, usually requiring simpler hardware, where a PHT is then used to modulate and enhance the same property. Most studies confirmed the effectiveness of combined multimodal approaches. However, in the research work [61], most participants reported that multiple feedback modalities were more confusing than a single modality. The study had participants combine pseudo-haptic visual light with sound and vibration to perform a robot arm's remote grasping manipulation of an object. Most participants found the feedback from multiple modalities confusing compared to that from a single modality and considered the sound feedback noisy. The authors concluded that efficiency slightly decreased, the operator's cognitive load increased, and the combined multimodal interface performed worse than any single feedback modality.

Therefore, PHT design requires understanding the correlation between different factors and multimodality dependencies. For example, in [71], results indicate that, although the displacement technique perceives the widely used stiffness effect, the object color can interfere with its effectiveness. Specifically, the predominantly yellow color of a sponge material made it more challenging to visualize the effect. The study [75] concluded that vibrotactile stimulation on fingertips causes a sensation of elbow joint displacement only when the pseudo-haptic effect does not prevail over the haptic illusion effect. Or, when a ball has a high volume and significant presence in the visual space, participants rely more on pseudo-haptics visual cues than force feedback. Nevertheless, most studies, like [75], conclude that pseudo-haptics combined with different haptic modalities produce more accurate responses than haptics alone.

### A. HAPTIC CUES

Simple passive haptic cues were present in the surveyed articles presented in Table 1 (16, 19, 22, 24) and Table 2 (5, 8, 9, 11, 12, 14, 22, 23, 24, 26, 28, 29, 32). These cues can facilitate an interactive experience by giving the user the sensation of physically grasping something. However, more elaborate haptic cues were also explored in the surveyed articles.

A combined PHT input system for walking-in-place (WIP), based on elastic passive strips and by adjusting the output speed gain to the stepping gesture, simulated the load sensations when walking on a slope. The PHT multimodal approach showed an improvement in realism with the virtual environment [76]. Similar to previous research on virtual spring stiffness [6], the work [77] compared the stiffness of real and pseudo-haptic springs in discrimination tests. In the study, pseudo-haptic stiffness varied between −40% and 60%, from the targeted stiffness, since subjects' perceptions were no longer influenced by visual feedback outside this range. This experiment helps to understand how pseudo-haptic effects are perceived under combined scenarios with real physical cues. Interestingly, the force perceived on the pseudo-haptic spring was greater than that on the real spring for the same stiffness. The results suggest that subjects underestimated the stiffness of the pseudo-haptic spring. This finding could be relevant in other contexts, such as rendering different stiffness levels in biological tissues.

The study [34] demonstrated how PHT can provide kinesthetic feedback to users. They used a common VR tangible cue and pseudo-haptics visual manipulation displacement where the object's resistance determines the offset. When pseudo-haptics are off, if an object cannot be moved, the virtual hand follows the tracked one, penetrating the object. When an object can be moved, there is no representation of physical resistance. In contrast, when using pseudo-haptic feedback, the virtual hand does not penetrate the immovable object, and the offset between the tangible cue tracker and the virtual hand increases. The object resistance is perceived as no longer following the exact tracking position (the offset depends on the object's physical resistance). According to different experiments in the same study, combining pseudo-haptic effects with vibrations used to induce kinesthetic feedback, pseudo-haptic effects were the predominant influence. In contrast, vibration served as a supportive channel for visual effects. This multimodal combination enhances the immersive user experience. The visual pseudo-haptic combined with vibrotactile is one of the most common multimodality combinations and will be described in more detail in the next section.

### B. VIBRATORY AND SQUEEZE

As described before, the lack of feedback when there is no physical contact, such as between a cursor manipulation and its targets, can be overcome by incorporating a haptic device within a controller to provide tactile information when a virtual object is touched. These devices can also be combined with visual PHT in a multimodal approach. This is evident in the surveyed articles in Table 1 (4, 6, 8, 15, 17, 21, 22, 27) and Table 2 (15, 25). Combining PHT with vibratory or squeezable haptic devices is one of the most commonly used configurations in experiments. This suggests that PHT combined with vibratory haptic modalities produces more accurate responses than either in isolation. For instance, works [75] and [78] demonstrated rendering virtual objects' stiffness by combining pseudo-haptic visual displacement with a handheld vibrotactile controller, applying squeeze forces at the fingertips.

It was shown that vibrotactile roughness can be modulated by combining visual changes of the display cursor with a vibrotactile stylus [40]. Using pseudo-haptics visual perturbations, users perceived the touched surface to be approximately 80% rougher, significantly altering their tactile experience. In [79] the authors propose a squeeze and vibrotactile wrist device for haptic feedback in various AR/VR interactions. They combine this device with pseudo-haptic feedback to simulate the hardness properties of a virtual push button. By adjusting the C/D ratio displacement between the user's real and virtual hands, they achieve different hardness levels. The interaction begins with vibrotactile feedback when the user's avatar finger touches the button's surface, followed by a proportional squeeze to convey the intended stiffness as the user presses the button. Users perceive different stiffness levels through pseudo-haptic (C/D) ratio control, where a stiffer button requires increased real-world movement. The results show that this multimodal approach produces more accurate interpretations and responses to virtual button stiffness than using either modality alone [79].

The work in [80] further improved interactive experiences via multimodal configurations. Researchers combined two PHTs: visual soft surface deformation, and cursor speed, extending their technique to create a multi-point PHT. They explored virtual surface interaction on a touchscreen with three-finger contact points and compared nodule detection sensitivity in tissue. They conducted experiments with vibration motor feedback indicating proximity to a hard nodule when moving from a softer to a stiffer region. Results showed multi-finger interactions were faster than single-point PHTs and performed similarly to vibration-based feedback for nodule detection sensitivity. The combined feedback technique was the most effective, with 71.4% of participants preferring it.

Another study [71], evaluated stiffness perception in smartphone touchscreen interfaces, simulating interactions with different materials. Participants pressed six materials (rubber, sponge, fabric, plastic, cardboard, and wood) and corresponding virtual materials. The study included the smartphone's native vibration. Results indicated participants perceived the pseudo-haptic stiffness effect in most materials similarly to actual materials. Virtual materials showed only a slight increase in perceived stiffness compared to real materials. The combination of vibration and pseudo-haptics enhanced and mitigated perceived differences between real and virtual materials due to the inflexible smartphone display.

Another study combined different haptic interfaces with visual representation in mid-air images. A game [57] integrated CG characters' movements, displayed with mid-air imaging, with indirect haptic feedback via a vibrotactile object controlled by a small robot. The experiments involved pushing or pulling movements on the CG character. For instance, moving the CG character by pushing the small robot while it shrinks, where the shrinking representation added force in the reference, and moving the CG character by hitting the robot while rotating. The results showed that pseudo-haptic feedback effectively changed the perceived direction of the force, despite the vibration from the vibrotactile cue being the same. These findings suggest that cross-modality, based on combining pseudo-haptics with active haptic device actuators, improves intuitiveness, attractiveness, and immersiveness [27].

### C. HUMAN BODY

The surveyed articles presented in Table 1 (5, 11, 14, 26) and Table 2 (2, 4, 11, 13, 16, 18, 27, 30 to 33) explored complementary visual PHT with the proprioception of the human body as an enhanced multimodal experience. For example, [59] found that pseudo-haptic visual feedback helps convey the sensation of grasping a virtual object but does not help maintain the grasp. Instead, subjects found it easier to maintain a virtual object grasp when their fingertips were in contact, as proposed in the "Self-Contact" method. Their study concluded that integrating human body fingers into the method was more effective. Another approach combined muscle tension with PHT. Simulating virtual force by combining muscle tension with additional pseudo-haptic feedback [47]. The displacement offset between the real hand and its virtual representation depends on muscle tension. When users flex their biceps, they reduce the offset, resulting in higher muscle force in the virtual world. Heavier objects with higher inertia and friction require more significant displacements. This technique combined with muscle tension input feedback allows lifting heavier objects with lower displacement depending on muscle flexing tension. Inner tendon vibration while lifting a virtual object extends the PHT motion range gain, without users detecting visual physical discrepancy by about 13% [81]. An intimate heartbeat can evoke a sense of virtual touch, through multimodal PHT visual heartbeat oscillating in size and color, with the sound of a realistic heartbeat increasing in volume as it approaches [22].

Other research areas explore combining PHT with human body modalities. Notably, the combined pseudo-haptics C/D ratio and electrical muscle stimulation (EMS) to manipulate virtual weight perception [45]. EMS stimuli cause triceps contraction and biceps relaxation, making it feel more difficult to lift a virtual object. This combined approach increased the modulation range of simulated weight by about 2.5 times. However, these methods see limited long-term usage due to fatigue and discomfort. Participants felt uncomfortable or unfamiliar with EMS-induced sensations [45]. EMS can magnify the effect when required, such as for lower C/D ratios, only PHT suffices. The EMS approach advantages include smaller size, lower battery, and power consumption, and eliminating the need for heavier robotic exoskeletons. The combined approach carries advantages, mitigating challenges by reducing the required EMS intensity. EMS proved beneficial for simulated weight perception when the visual effect was ambiguous. As future work, one could consider leveraging the EMS research by Pedro Lopes et al. [82], [83], [84] and [85] combined with PHT potential benefits, which from literature review seems not to be explored so far.

*D. ACOUSTIC*

As presented in Table 1 (1, 19, 23, 26, 27) and Table 2 (1, 25), those surveyed articles proposed a multimodal approach with acoustic feedback. Various PHT studies used simultaneous simulation effects with good results. The work [73] presented a multimodal approach combining pseudo-haptics with a visual cursor and a vibratory mouse to simulate roughness, hardness, and friction with acoustic sensory stimulation in a collocated touchscreen texture exploration task via a pen-guided path. The results indicate that auditory and vibratory feedback are equally effective in enhancing the task, supporting the proposed approach. However, when the path was visually shown, neither pseudo-haptics image deformations nor auditory and vibratory feedback induced significant behavioral changes because the visual dominance sufficed to perform the task. The task was about 50% slower without visual path information.

Another study investigated the effect of delay in auditory feedback on the sense of heaviness [86]. The results suggest that frequency contributes more to the heaviness sensation than loudness. For example, pseudo-haptic sound stimuli can be applied in AR when people touch, grasp, or release a virtual object directly with their hands in a mid-air VR HMD environment. In [87], participants felt the virtual object was softer and more deformable when the volume increase was emphasized. A related analysis concluded that using auditory stimuli instead of tactile feedback can provide the accuracy required for surgical operations where direct touch is impossible [88]. However, other studies such as "HapticHead" [89], found that AR/VR using multiple vibrotactile actuators around the head helps users find virtual objects more quickly and accurately with vibrotactile feedback. Although sound-based techniques can be efficient, some participants described the sound feedback modality as noisy [61]. This perception was supported by measurements of the operator's brain flow, showing the highest information processing load with sound feedback.

Apart from auditory effects, other acoustic applications have been studied, such as ultrasound applications [23]. For instance, acoustic levitation displays use an array of ultrasound to produce acoustic pressure patterns, levitating small physical particles, typically tiny polystyrene beads, in mid-air. The emitted ultrasound standing waves are localized above and below the particles, to display complex object structures. This technology, akin to a hologram, enables contactless gesture interactions.. Pseudo-haptic effects can enhance this display, simulating touch feedback like surface friction, sliding, texture, stickiness, or an object's mass and inertia. For instance, in a friction simulation, a user slides particles positioned in a virtual cube mid-air over a predefined surface area. Friction sense is simulated by dynamically reducing velocity (C/D ratio control) and creating a discrepancy between gesture movement amplitude and particle displacement distance. Greater displacement discrepancy intensifies the friction sensation. This technique could guide users in dragging a levitated particle from point A to point B in space without optical path cues, using friction feedback to indicate path borders.

*E. ACTUATED DISPLAYS*

Actuated Display which include shape-changing displays can reconfigure depending on displayed content or task. They can for example, separate and join for collaborative displays or adjust peripheral devices according to user positions and activities. One study proposed the "BouncyScreen" technique, utilizing physical screen movements to simulate force feedback when users manipulate a virtual object with a VR controller [58] including three different use cases described in Table 1 (3) and Table 2 (3, 17). This work proposed a "BouncyScreen" force technique, achieved through the screen's physical movements when users manipulate a virtual object using a VR controller device. The study conducted various physical experiments to assess BouncyScreen's effectiveness when combined with different PHT methods. One approach involved fixing the virtual object with no pseudo-haptic effect, relying solely on physical screen movements. Another approach used a combined pseudo-haptic C/D ratio method to create visual effects synchronized with screen movements, simulating weight, hardness, and force. The results suggest that induced pseudo-force feedback through physical screen movement was almost as effective as vision-based pseudo-haptic feedback. Furthermore, additional studies revealed that BouncyScreen could represent different forces for distinct operations, such as pushing and bumping, significantly enhancing the sense of presence and contact during interaction.

*F. MULTIMODAL FORCE FEEDBACK DEVICES*

Some experiments suggest that pseudo-haptic stimuli can be effectively combined with force feedback. Previous studies reported that a force feedback device combined with pseudo-haptics provides a better hardness perception [90]. The experiments showed that a force of approximately 0.2 N – 0.4 N is perceived more accurately than with only real haptics. Additionally, the magnitude of the difference was nearly 10% of real haptic feedback. This specific multimodal combination suggests the potential for more compact haptic force-feedback devices. Another study demonstrated enhanced results by adapting the trigger resistance of the VR controller combined with the PHT C/D ratio during object lifting to create a weight sensation [91]. Participants were more sensitive to smaller weight differences in the combined weight simulations compared to the individual methods and were able to determine weight differences faster.

## VII. PSEUDO-HAPTICS APPLICATION USE CASES

This section addresses RQ5 by presenting the main pseudo-haptics applications in practical use cases. While most articles discuss the effectiveness, simplicity, and applicability of pseudo-haptics, few explicitly apply those to concrete, practical applications. However, 36% of the surveyed articles do present interesting application studies.

Pseudo-haptics can be applied in various contexts requiring haptic feedback across many types of displays, proving to be an effective and low-cost technique for simulating perceptions. The current review's taxonomic method evaluated the effect of pseudo-haptics in a concrete application context. Articles proposing potential applications without concrete implementation were not mapped in the application category. PHTs have been applied in various professional and entertainment industries, including education, training, sports, gaming, medical treatment, military, and manufacturing. Similarly to the categorization in [92], we propose grouping those into three categories for haptic systems in virtual environments: learning, assistance, and entertainment. These categories are described in the following sections.

*A. LEARNING APPLICATIONS*

Learning applications encompass tools or strategies designed to acquire new knowledge and competencies related to specific tasks or topics. This category includes applications in education, learning, and training. The surveyed articles [80], [15] and [60], focus on learning applications which are fewer compared to those on assistance or entertainment. This scarcity might stem from the challenge of aligning the simulation environment with real-world scenarios. Evaluating PHT in conjunction with other multimodal techniques could enhance the realism of training environments by providing a better sense of physical stimuli. Pseudo-haptics standalone feedback may be effective in scenarios involving smooth, prolonged movements, such as underwater environments, where the need for external physical devices is minimized.

Medical training and robotic surgical assistance are significant areas of interest for pseudo-haptics. However, many medical training systems remain inaccessible to hospitals and universities due to high costs and complexity in hardware and software design, configuration, and maintenance. These systems aim to provide medical students or novice surgeons with realistic experiences akin to real operating theaters but within controlled virtual environments [80], [15]. For instance, pseudo-haptics have been effectively used to simulate various virtual tissue characteristics for surgery, aiding in training for anesthetic procedures, tumor identification, and palpation necessary to locate arteries or organs under the skin by simulating the corresponding bumps and hollows. Pseudo-haptics conveyed stiffness information and tumor hardness on simulated soft tissue surfaces in virtual environments [80]. Inputs included computer mouse or touchscreen-sensitive devices, which effectively distinguished different sizes of virtual hard nodules integrated into the simulated soft bodies. Additionally, using a haptic feedback system [15] demonstrated the effectiveness of pseudo-haptics in robotic-assisted laparoscopic surgery (RALS) and its training.

*B. ASSISTANCE APPLICATIONS*

The following surveyed articles included assistance applications: [73] [61], [93], [35], [56], [94] , and [2].

**Teleoperation.** Teleoperation involves remotely controlling robot systems in various scenarios, such as in remote medical surgery, maintenance tasks, and inspections in inaccessible areas after natural disasters. It is crucial for executing complex tasks in hostile environments, such as space or underwater. Haptic feedback is essential for manipulating remote machines and superimposed pseudo-haptic visual feedback can reduce equipment cost and complexity. For instance, pseudo-haptics can enhance force sensing to help users track precise paths during manual handling tasks. Results show a 50% improvement in manual accuracy with pseudo-haptics, although the time to track a path increased with higher expansion rates [95].

In teleoperated surgical robotic systems for minimally invasive surgery, pseudo-haptics replace kinesthetic haptic feedback devices. The system provides pseudo-haptic hardness feedback based on the relationship between two forces: the user's grip force ($Fg$), on the sensor and the real material force measured at the end-effector ($F_e$). This setup reflects real force interactions rather than virtual material properties. The experiment demonstrated that visualizing the altered displacement of the end-effector ($x_e$), effectively provided the intended pseudo-haptic feedback [94] :

$$x_e = f(F_g, F_e) \qquad (2)$$

This PHT reduces the need for conventional actuators and kinematic apparatus thus avoiding the constraints and limitations of traditional haptic feedback. It also overcomes the limited degrees of freedom (DoF) typically associated with such systems. The master movement, constrained by its native DoFs, limits the DoF movements provided to the slave system in remote robot teleoperation assistance.

In a prototype humanoid remote machine, PHT was applied with superimposed visual image feedback on the robot's fingertips [61]. The operator receives haptic feedback when the robot arm grasps an object, as the pressure sensor attached to the tip of the gripper exceeds a threshold value. Experimental studies concluded that pseudo-haptic visual light feedback reduces grasping force by 24.1% and lowers the operator's cognitive load when using pseudo-haptic visible light with sound and vibration during remote object manipulation. Further evaluations of visual haptics implementations, where haptic information was superimposed on the object's point of contact, showed that visual haptics effectively facilitated force perception. A direct LED display mounted on the fingertip lit up according to the applied pressure, and computer vision techniques identified the fingertip's spatial position from a camera image. A haptic image was superimposed on the robot hand's fingertips according to the pressure sensor's value. When only visual-haptic feedback was used, results revealed that visual haptics stabilized the grasping and carrying performance of fragile objects, making them suitable for operable remote robot machines without complex and expensive interfaces.

**GUI Design**. Gamification techniques and PHT can enhance GUI applications. Early studies [16] described pseudo-haptics in GUI design, where a high C/D ratio allows faster navigation, while a low C/D ratio aids in fine adjustments. Varying C/D ratios enable pseudo-haptic effects like sticky, magnetic, or repulsive icons. In multi-display contexts, PHTs can create sticky widgets at the edge of a monitor, reducing accidental transfers and selection times. In manual data mining of digitized tissue slices, PHT combined with automatic zooming improves visual navigation accuracy and task completion time. Some users described a "magical force" guiding the cursor to the target position.

In this survey, we saw that PHT was also used to perform user interface navigation on touchscreens [93]. The proposed PHT captures user attention during interactive scrolling or browsing on touchscreens. Results showed that figures with higher frictional feedback were better retained in memory. Participants performed best in visual memory tasks using interactive scrolling with dynamic pseudo-haptic kinetic friction. This method leverages cognitive factors of short-term memory, relating user attention during browsing to memory retention. PHT can effectively capture user attention in GUI design, advertisements, mental care, or education applications. In a mid-air GUI interface [35], a 3D carousel ring menu with pseudo-haptic feedback can repulse or attract user attention to specific items as they navigate menu options using swipe gestures.

**Medical Rehabilitation and Workout**. The possibility of changing the perceived weight of objects via PHT can be applied to medical rehabilitation or workout activities. Studies have concluded that intensive motor training with VR games can effectively aid patients' arm rehabilitation by providing real-time performance feedback [56] This approach is less costly and labor-intensive than repetitive movement exercises performed one-on-one with a therapist. Participants controlled a cursor with their upper limb to perform a path-following task. The pseudo-haptics method adds motion assistance for path guidance and motion resistance to upper-limb rehabilitation in a VR environment. Results show that the motion assistance mode better tolerates trajectory deviations and is more suitable for early rehabilitation stages. It is also more time-efficient and accessible than motion resistance mode, which is more effective in later stages for muscle strengthening, endurance training, and fine motor control. VR avatar body movement illusions were also applied to stroke rehabilitation, requiring high body ownership by applying unnoticeable offsets [63].

## C. ENTERTAINMENT APPLICATIONS

Entertainment developers focus on gaming applications to create more immersive and enjoyable user experiences in virtual environments. Commercially available video games have leveraged optical illusions (which trick eyes into seeing something that is not really there[1]). For example, Sony Echochrome (2008) [2], pioneered integrating visual illusions in gameplay where a character navigates a world where physics and reality depend on perspective. Similarly, Monument Valley (2014) [3] and Superliminal (2019) [4] are first-person puzzle games based on forced perspective and optical illusions. To enhance gaming experiences, a visual illusion effects database has been developed for integration into video games, applicable to objects' appearances and intended perceptual effects [96]. Optical illusion and pseudo-haptic effects based on the C/D control technique have been combined in gaming, such as driving simulators. Thus, although the user manipulates the control input in the same way, the vehicle in the virtual environment can become more challenging to control when driving over slippery areas (e.g., oil on the road), slowing down on off-road sandy surfaces, or becoming heavier during the game action.

Among the surveyed articles, the following focus on entertainment applications: [57], [47], [58], [3], [46], [34],

---

[1] Cambridge Online: https://dictionary.cambridge.org/us/dictionary/english-portuguese/optical-illusion?q=optical+illusions. Accessed 26 Mar 2022

[2] Echochrome Sony: https://www.sony.co.in/microsite/playstation/product/echochrome/game.html. Accessed 26 Mar 2022.

[3] Monument Valley: https://www.monumentvalleygame.com/mv2. Accessed 26 Mar 2022

[4] Superliminal: https://store.steampowered.com/app/1049410/Superliminal/. Accessed 26 Mar 2022.

[63], and [22]. Those studies demonstrated that pseudo-haptics enhance user presence, immersion, and enjoyment in entertainment and gaming experiences. For example "BouncyScreen" [58] employs physical screen movements while the user manipulates virtual objects with a VR controller. Their study prototyped concrete applications, including a baseball pitching game where a character catches a ball thrown by the user. Users can perceive the force and speed with which they throw the ball by adjusting the C/D ratio and the speed of the "BouncyScreen", which moves backward in synch with the catch.

Another study simulated ball weight in a virtual bowling game [46]. Gaming realism improved when allowing a stronger game character to lift heavier objects, with less muscle tension and pseudo-haptics displacement. When a player becomes fatigued, the virtual world action requires more effort [47]. Kang et al. [57] proposed a game system based on the visual representation of CGC movement, displayed in mid-air imaging technology, combined with vibrotactile haptic feedback generated by a robot controlling the CGC movements. Another study [3] reproduced drag forces in a virtual underwater environment using pseudo-haptics where the player seeks to approach and touch dynamic sea creatures. The results show that the method effectively reproduced a realistic immersion in water entertainment activities, making slow limb motion a natural outcome of being underwater. This held even for participants with previous aquatic experience, such as swimming or diving. The study also identified potential applications for the early phases of scuba diving training.

## VIII. CHALLENGES AND LIMITATIONS
This section describes the challenges and limitations of the surveyed PHTs, and possible workarounds.

### A. VISUAL FOCUS
Pseudo-haptics relying on visual illusions, face the constraint that users must directly look at and focus on the specific area of interaction to perceive the intended effects [29]. Unlike other sensory stimuli (e.g., audio, or haptic device actuation), which can be perceived even without direct focus, pseudo-haptic visual stimuli require the users' full attention. However, situations requiring focused attention on a critical visualization area can benefit from this limitation. Eye-tracking systems could also detect when the user is not paying attention to the required focus area, potentially mitigating it.

### B. VISUALIZATION MISMATCH
The displacement technique is the most common in pseudo-haptics, enabling various sensory simulations. However, this method can lead to mismatches between the user's actual input and the visual stimuli, such as positional or rotational discrepancies. These mismatches are problematic in environments where user input is visible. For instance, in pseudo-haptic weight rendering, increased displacement intensifies the perceived weight but can lead to unnatural feedback or a loss of agency if excessive.

Different approaches have been proposed to address these challenges. For example, [97] suggests beyond-real VR interactions, that intentionally include sensory mismatches to remap spacetime or alter user representation. Another study [46] accepted mismatches as part of the design, using noticeable offsets between tracked hands or VR controllers and their virtual positions to allow a broader range of perceivable weights. Participants in a bowling game experiment comfortably accepted up to 24 cm displacement offsets for weight mapping distortion. However, touchscreens and see-through HMDs, where input and visual stimuli are co-visualized, make noticeable mismatches more apparent, complicating pseudo-haptics induction.

Next, this section will discuss techniques from the survey articles aimed at overcoming these challenges across different visualization media.

#### 1) MID-AIR BODY DISPLACEMENT MISMATCH
Experimental results showed that faster hand movements in mid-air tracking situations, led to a loss of virtual force perception created by pseudo-haptic feedback, whereas slower movements maintained the perception [56]. Smaller discrepancies in pseudo-haptics are more accurately perceived when people slow down during tasks requiring precision, suggesting PHT's suitability, for slow-movement applications, like arm rehabilitation therapy.

To address visualization mismatch displacement, some articles propose various workarounds. For instance, a decreased sense of presence occurs when actual body motion significantly deviates from the avatar's virtual motion, preventing natural limb movements [3]. To mitigate this, a design guideline suggests defining a maximum allowable deviation or gradually decreasing the deviation over time. Additionally, a "World Warping" technique manipulates the virtual worldview without the user's awareness, correcting the view when the user moves their head, performs a task, or blinks [67].

#### 2) TOUCHSCREEN DISPLACEMENT MISMATCH
Pseudo-haptic effects can enhance touchscreen perception, but pose a design challenge where real and virtual fingers coexist and move to different positions. Different approaches have been proposed to address this issue [37], [7], [80]. One method separates the user input area from the display visualization zone, ensuring there is no visible mismatch, as users do not directly see their fingers. Another approach uses background images as holistic visual feedback, instead of focusing on a specific small display zone. Workarounds include applying larger visual stimuli, allowing the actual user body to remain within the visual stimuli range. For instance, scrolling a background image with various distorted displacement ratios between the user's finger and the background induces frictional sensations. This method ensures that regardless of the ratio, the user's finger on the touchscreen touches the background image, avoiding mismatches. Using a large cursor size on the touchscreen also maintains the illusion of contact between the finger and the cursor, even with significant displacement differences.

Another workaround uses a virtual visual string connecting the finger and the object. When users touch and swipe the object, they notice a shift due to the configured displacement ratio. The string appears when there is an effective displacement between the input finger's position and the object and disappears when the distance exceeds a "break effect" threshold or the finger is lifted. Where the virtual string effectively evokes pseudo-haptics.

### C. SIMULTANEOUS SIMULATION

According to our taxonomy, despite the variety of pseudo-haptics simulation techniques, evaluations mainly focus on a single effect. Simultaneous usage of different PHTs may raise challenges, though some studies explore this.

#### 1) HUMAN BODY SIMULTANEOUS SENSE SIMULATION

The study [80] examined two simultaneous surface stiffness display PHTs: surface deformation and cursor speed modification. In multi-finger interactions, three identical avatars move uniformly in the x and y directions, with each finger translated independently in the z-direction according to its stiffness value. This allows examining the surface stiffness of a neighboring zone. Evaluating how simultaneous pseudo-haptics are perceived when different PHTs are used together is intriguing. Research opportunities include studying PHT relationships, interferences, and the challenges of designing and controlling complex approaches. For example, most interfaces for weight perception use one hand, yet real-life scenarios often involve two hands. A load feels lighter when lifted with both hands, so the applied force should decrease [2]. This bimanual study area is relevant for carrying large, heavy objects, and enhancing usability in XR environments featuring bimanual interactions and immersive multi-sense experiences. The scarcity of studies in this direction warrants further research to increase the method's effectiveness and immersion.

#### 2) MULTIMODALITY CHALLENGES

Regarding multimodal sense perception, one must consider potential sense conflicts. Simultaneous stimulation of different senses or different haptic cues can occur. E.g., physical proxy shape, mass, and mass distribution can concur with PHT to generate weight perception during object rotation [48]. This conflict can elicit design opportunities in PHT, where visual stimuli typically prevail. However, multiple senses and actuators acting simultaneously create challenges. Future research should explore relationships between different manipulation techniques and pseudo-haptic experiences, such as controller-based interactions for full-body avatars, and how the manipulated physical body part affects the pseudo-haptic experience [49].

### D. SIMULATION VERSUS REAL

Although pseudo-haptics can enhance sensory perception, it is crucial to remember that these remain simulations. There is no direct interaction with tangible objects or natural gravitational forces. Users often perceive interfaces as lacking real-world plausibility. Indeed, only a few studies have mapped the perception of weight to quantitative results, as in [2]. The work [46] suggested a technique based on perceived relative weight differences using various distortion offsets, rather than mapping to an absolute weight value. This offset approach is more effective for comparing two objects' weights than for determining a single object's weight. A quantified value of illusory weight is presented for a haptic cue reference, noting that pseudo-haptic feedback can overestimate or underestimate the actual weight [44]. Future research should investigate quantitative analysis methods for perceived absolute values of mass and mass difference to improve the accuracy of perception with pseudo-haptics, confirming their potential advantage over haptic devices in rendering objects' mass.

In surface features, an experimental study directly mapped the perceived effects of virtual sense simulation to natural material surface characteristics [18]. For each material, participants indicated the virtual descriptor closest to the perceived simulated effect. Results suggest that the proposed PHT was adequate for inducing different haptic sensations, with no effect along other dimensions.

This challenge of simulated versus real perception is also present in medical training applications [80]. Pseudo-haptics are applied to simulate remotely palpated physical characteristics of human tissue for training purposes. PHT training systems allow users to virtually explore patients' bodies, providing a realistic and controlled environment for surgical training. Further studies are needed to evaluate the effectiveness of virtual training in real scenarios. Conclusions from tumor identification during palpation suggest that pseudo-haptics provide adequate stiffness perception for soft objects, with high fidelity for hard inclusions within soft tissues in surgical training systems. However, the effect on real tumor identification performance post-training remains unclear and is a target for future study. While certain real-case scenarios were addressed in other work [15], significant evidence supports using pseudo-haptic feedback in RALS systems and training. Future research should explore more complex medical tasks, such as suturing and needle insertion, and assess how pseudo-haptics could extend to more complex real-world scenarios. Further studies are essential to confirm the effectiveness of virtual environment training in real-world applications, not only in the medical field but also in other areas.

### E. COGNITIVE FACTORS

Cognitive factors are another aspect that must be considered in pseudo-haptics design.

#### 1) SENSORIMOTOR MEMORY BIAS FACTORS

Understanding cognitive sensorimotor memory bias factors can be illustrated through weight perception experiments [5]:

- Object lifting pre-experience results in force scaling.

- Users lifted dolls of the same physical weight but representing different human individuals (age, sex), more carefully, biased by the social cue expecting females to weigh less than males.

- Golfers and non-golfers were presented with actual golf balls and practice balls modified to weigh the same. While non-golfers judged them to weigh the same, golfers, with previous knowledge, expected the practice balls to be lighter, leading to a misperception of the weight.

#### 2) GUIDE AND ATTRACT THE USER'S ATTENTION

Pseudo-haptics can guide and attract user attention in GUI applications by leveraging cognitive factors. In one study, researchers explored methods to draw users' attention to specific display zones [33]. Traditional methods involved modifying the color, contrast, or resolution of regions, but these required altering the original content. Other techniques redirected the user's virtual camera to shift closer to a target, akin to "snapping" with features like "magnetism" or "gravity." These methods influence user interest, aiding learning and memorization due to the interplay between perception and cognition, impacting attention and preferences. For instance, rough objects complicate social interactions, while hard objects increase negotiation rigidity. By introducing pseudo-haptic resistive feedback in areas requiring more cognitive effort, users can be encouraged to pay more attention and perceive the information as more critical, facilitating learning and retention.

#### 3) COGNITIVE LEARNING PROCESS

In ecological psychology studies [99], typical HCI involves users performing tasks based on the physical meaning of displayed symbols, representing a complex environment through rational deductive reasoning. Rapid reaction and improvisation require direct control over the process, necessitating "transparent" display systems. The concept of transparency, introduced by "Ecological Interface Design" (EID) [100], replaces the rote instrument approach (single-sensor-single-indicator), which is more suitable for robot instructions, than human use. EID was initially devised for industrial displays, such as those used in nuclear power plant control rooms, and aligns with ecological psychology principles, providing a comprehensive account of human behavior akin to cognitive engineering [100]. By supporting human capabilities, perceptual mechanisms, and planned actions, EID facilitates the learning process and human problem-solving, essential for building more natural XR systems.

Another study [101] suggested age and gender differences in the PHT perception, with younger or male participants experiencing greater effects. These findings appear related to users' daily habituation of mouse interfaces and the accuracy of detecting pointer positions through vision or proprioception. This study highlights the importance of considering user diversity in HCI design, suggesting that interface calibration should account for habituation rather than age or gender.

The role of memory in the pseudo-haptics cognitive learning process was emphasized in [94]. Experimental results indicated that sleep plays a crucial role in memorizing sensory impressions of pseudo-haptic feedback. Days-long training periods may be necessary for operators to master teleoperation systems' learning curves and adapt to their use.

#### 4) COGNITIVE WORKLOAD

Personnel performing teleoperation tasks must manage a high cognitive workload in Situational Awareness activities, by continuously perceiving relevant elements, understanding their significance, and projecting required near-future actions based on status assessment [102]. NASA studies [103], show that task performance is significantly reduced when the cognitive workload is too high, situation awareness is lost, or when operators are overly stressed or fatigued.

Regarding PHT, a method to evaluate and compare different multimodal efficiencies was proposed based on the required average grasping force and cognitive load measured by electroencephalogram (EEG) [61]. In this study, density arrows indicated the direction of information flow in the brain, with arrow thickness representing the flow and numerical values indicating total information flow for each modality. The EEG results found that visual feedback (superimposing haptic information as images on the robot's contact points) was most effective, reducing the grasping force by 24.1% without increasing brain information flow. This pseudo-haptic visual light feedback was more efficient than sound and vibration feedback in controlling grasping force and had a lower cognitive load. This lower cognitive load is likely due to operators focusing solely on the remote robot hand, where haptic information was conveyed visually, eliminating the need to switch focus. Therefore, PHT can be considered a cognitive aid, especially visual feedback focused on the operation zone, assisting tasks instead of a knowledgeable controller or intelligent decision support system [103].

For human sensing and performance monitoring, real-time physiological workload measurements seems to be more suitable [103]. EEG measurements provide a practical, objective, and quantitative evaluation of methods and could be used for general task performance and cognitive load assessment. Additionally, functional near-infrared spectroscopy (fNIRS) is commonly used to assess cognitive workload [102] and could potentially be used for teleoperation cognitive load assessment, alongside less intrusive sensors like cardiac activity, eye blink rate, and electrodermal activity (EDA). These can assess changes in neurophysiology, physiology, and behavior (e.g., fatigue, overload, attention, engagement, and expectancy), providing real-time, continuous, and objective physiological measurements.

## IX. FUTURE RESEARCH OPPORTUNITIES

Based on our survey findings, we propose new areas for future research, focusing on combining the advantages of PHT in multimodal applications. These areas aim to enhance user sensations and perceptions through pseudo-haptics, integrated with more diverse and natural physical stimuli.

**Body Ownership and Kinesthetic Illusion**. Future research should focus on body-ownership and kinesthetic illusions aiming to understand multisensory integration in virtual environments better. This includes investigating vision and touch combinations, and synchronizing haptic and visual cues.

**EMS and PHT Multimodalities**. One promising future research area involves combining PHT with EMS applications [82], [83], [84] and [85]. Although those previous work did not use PHT, combining it with EMS could offer advantages, such as reducing the intensity of EMS required, thus minimizing human stimulus, and device power consumption. Also, when EMS interacts with the human body (in assistance or guidance use cases), maintaining the user's sense of agency is crucial. research could explore whether combining EMS with PHT methods helps preserve the user's sense of agency, by providing early visual cues to improve situation awareness, avoiding a standalone EMS action.

**Underwater VR and Robotic Teleoperation**. Slow movements enhance the perception of the virtual force created by PHT [58], [56]. In underwater environments, natural physics causes movements to be slower making PHT even more suitable for these simulations. For example, robotic teleoperation with remotely operated vehicles (ROVs) in aquatic conditions requires subtler discrepancy illusions to simulate motions properly. PHT applications in robotic underwater teleoperation for situation awareness have not been addressed in the surveyed papers. This area, along with other domains, presents exciting opportunities for future research.

**HCI design and XR affordance tasks**. Future research work should focus on XR affordance task design, using hybrid approaches where PHT is further explored and applied in multimodal approaches. HCI design could evolve to incorporate XR affordance tasks in the context of Ecological Interface Design (EID). This would allow users to subconsciously break down information patterns in novel situations into familiar affordance perceptions, leveraging PHT. The goal is to provide an affordance perception of the work domain through the interface analogous to human capabilities, perceptual mechanisms, and actions, supporting problem-solving.

## X. CONCLUSION

Our pseudo-haptics survey, assessed, summarized, and organized individual articles' findings into a comprehensive categorization based on a proposed taxonomy including tactile, kinesthetic, and composite feedback categories. The survey analyzed pseudo-haptics methods in terms of visual stimulus, the pseudo-haptic technique, the corresponding user actions, and visualization media. Our literature survey showed that pseudo-haptics can provide cost-effective and flexible haptic feedback, with potential applications in learning, assistance, and entertainment. The challenges and limitations of adopting pseudo-haptics methods were also evaluated, including the need for user attention on the visible display, the possibility of visual displacement mismatch, and simultaneous simulation challenges. However, these issues can be mitigated using techniques, such as virtual visual cues or incorporating displacement into the design process, among others. According to surveyed articles, the most common visualization media in pseudo-haptics are HMD devices, followed by touchscreen, and desktop displays. The survey identified a trend of increased use of pseudo-haptics in mir-air interactions, mapping visual effects to users' physical movements in XR environments, aligning with ecological psychology studies, that consider multimodal affordance perception. The survey showed the effectiveness of pseudo-haptics in multimodal contexts, with other modalities like vibratory/squeeze haptics cues and body movements, contributing to a more immersive experience. Finally, our survey identified potential future research opportunities for pseudo-haptics, such as exploring combined multimodal use cases in XR environments and enhancing robotic remote teleoperation. These insights can usher in new dimensions in user interaction and sensory experiences, heralding an exciting future for pseudo-haptics technology.

**Rui Xavier** is a PhD candidate of the doctoral Programme in Electrical and Computer Engineering (PDEEC), Instituto Superior Tecnico ("IST"), Universidade de Lisboa, with a degree in Electronic Systems and Computers Engineering also from IST. Industry professional experience is related to telecommunications, medical equipment, and electronics. Current main research interests are related with computer vision, machine learning, human-computer interaction and human-robot interaction, aiming to contribute for a multimodal immersive teleoperation and data visualization interfaces for improved awareness in ocean and space robotic exploration.

**José Luís Silva** received a joint Ph.D. degree in computer science from the Portuguese MAP-i Consortium (University of Minho, University of Aveiro, and University of Porto) in 2012. He is an Assistant Professor with the Department of Information Science and Technology, Lisbon University Institute (ISCTE-IUL), a senior researcher of the Interactive Technologies Institute (ITI) research unit of the Laboratory of Robotics and Engineering Systems (LARSyS), and a member of ACM Europe Technology Policy Committee, and IFIP TC 13 - Working Groups 13.2 and 13.10. Prof. Silva co-leads the MEROP research team, has participated in several international and national research projects (including with Airbus and the European Space Agency), and in the Portuguese contribution to Simulated Mars Missions: AMADEE-20 and AMADEE-24 organized by the Austrian Space Forum. His work has been published in prestigious peer-reviewed international journals and conferences such as IEEE Access, IJHCS, ACM EICS, IEEE RO-MAN, ACM IUI, ACM/IEEE HRI, and INTERACT. His main research interests lie in Interactive Systems, Human-Robot Interaction, and Space Exploration. His main awards and honors include ISCTE-IUL Scientific Awards and Ph.D. Award from Fraunhofer Portugal Challenge.

**Rodrigo Ventura** (PhD) is a tenured Associate Professor of the Electrical and Computer Engineering Department of Instituto Superior Técnico (IST), University of Lisbon, and a senior researcher of the Institute for Systems and Robotics (ISR-Lisbon), part of the Laboratory of Robotics and Engineering Systems (LARSyS). He is the coordinator of the Minor in Space Sciences and Technologies at IST. He collaborates regularly with the International Space University (ISU) in academic activities. Broadly, his research is focused on the intersection between Robotics and Artificial Intelligence, with particular interest in human-robot interaction, mobile manipulation, biologically inspired cognitive architectures, and machine learning. This research is driven by applications in space robotics, aerial robots, and service robots. His research in space robotics led to robotic experiments on the International Space Station (ISS) and in various analog space exploration missions.

**Joaquim Armando Pires Jorge**, a Full Professor at Instituto Superior Técnico da Universidade de Lisboa (UL), Portugal, teaching Virtual and Augmented Reality, Computer Graphics, and Human-Computer Interaction. He earned his PhD in Computer Science from Rensselaer Polytechnic Institute (NY) and leads the Graphics & Interaction Research Group at INESC-ID, affiliated with UL. Prof. Jorge holds the UNESCO Chair on Artificial Intelligence and Extended Reality. He has authored or co-authored over 380 peer-reviewed publications, including 74 journal articles and four Springer/Nature books. Prof. Jorge is also the Editor-in-Chief of the Computers and Graphics Journal (Elsevier). As an IEEE Senior Member, Distinguished Visitor and Contributor, he serves on the IEEE/CS Board of Governors (2023-2025). He is also Adjunct Professor at Victoria University Wellington, and Honorary Invited Professor at the Catholic University of Rio de Janeiro. He is a Fellow of the Eurographics Association, a Distinguished Member and Speaker of the Association for Computing Machinery, and a *Membre Libre* of the *Académie Nationale de Chirurgie* (French National Academy of Surgery).